\documentclass[useAMS,usenatbib]{mn2e}
\pdfoutput=1
\usepackage{natbib}
\usepackage{caption}
\usepackage{times}
\usepackage{verbatim}
\usepackage{amsmath}
\usepackage{amssymb}
\usepackage{graphicx}
\usepackage{ulem}
\usepackage{color}
\usepackage{rotating}
\usepackage{lscape}
\usepackage{aas_macros}
\usepackage{upgreek}
\usepackage{pdflscape}
\usepackage{afterpage}

\title[KINETyS: Constraining spatial variations of the stellar initial mass function in early-type galaxies]{KINETyS: Constraining spatial variations of the stellar initial mass function in early-type galaxies \thanks{Based on observations obtained at the Very Large Telescope of the European Southern Observatory. Programme IDs: 093.B-0693 and 094.B-0061}}

\author[P. D. Alton, R. J. Smith, and J. R. Lucey]{Padraig D. Alton$^{1}$\thanks{email: padraig.alton@durham.ac.uk}, Russell J. Smith$^{1}$, and John R. Lucey$^{1}$\\
	$^{1}$Centre for Extragalactic Astronomy, Department of Physics, Durham University, South Road, Durham DH1 3LE, UK}

\begin{document}
	
	\date{Submitted \today}
	
	\maketitle\label{firstpage}
	
	\newcommand{\rev}[1]{\textcolor{red}{#1}}
	
	\newcommand{\err}[2]{$\substack{+#1 \\ -#2}$}
	
	\newcommand{\gap}{\phantom{--}}
	
	\begin{abstract}
		
		The heavyweight stellar initial mass function (IMF) observed in the cores of massive early-type galaxies (ETGs) has been linked to formation of their cores in an initial swiftly-quenched rapid starburst. However, the outskirts of ETGs are thought to be assembled via the slow accumulation of smaller systems in which the star formation is less extreme; this suggests the form of the IMF should exhibit a radial trend in ETGs. Here we report radial stellar population gradients out to the half-light radii of a sample of eight nearby ETGs. Spatially resolved spectroscopy at 0.8--1.35$\umu$m from the VLT's KMOS instrument was used to measure radial trends in the strengths of a variety of IMF-sensitive absorption features (including some which are previously unexplored). We find weak or no radial variation in some of these which, given a radial IMF trend, ought to vary measurably, e.g. for the Wing-Ford band we measure a gradient of +0.06 $\pm$ 0.04 per decade in radius. 
		
		Using stellar population models to fit stacked and individual spectra, we infer that the measured radial changes in absorption feature strengths are primarily accounted for by abundance gradients which are fairly consistent across our sample (e.g. we derive an average [Na/H] gradient of --0.53$\pm$0.07). The inferred contribution of dwarf stars to the total light typically corresponds to a bottom heavy IMF, but we find no evidence for radial IMF variations in the majority of our sample galaxies.
		
	\end{abstract}
	
	
	\section{Introduction}
	
	The stellar initial mass function (IMF) is a key property of galaxies, describing the distribution of stellar masses found in a newly formed population of stars: understanding this distribution is invaluable for probing the physics of star formation. The assumptions made about the IMF's functional form have wide-ranging implications. The IMF determines the supernova rate and the chemical evolution of any population of stars and affects the derivation of key observables. These include the calibration of the star formation rate and, perhaps most importantly, the mass-to-light (M/L) ratio. Empirical estimates of galaxy stellar masses are therefore IMF-dependent.
	
	In our own Milky-Way galaxy the IMF is well-constrained through stellar censuses. It is well-described by a simple power-law from high masses down to 1M$_{\odot}$ \citep{1955ApJ...121..161S}. However, counts of faint stars have shown that this power-law breaks and turns over at sub-solar stellar masses (e.g. \citealp{2001MNRAS.322..231K}; \citealp{2003PASP..115..763C}). The IMF appears to be stable across a wide variety of Milky-Way environments, its form independent of local variations in e.g. metallicity, environment (as reviewed by \citealp{2010ARA&A..48..339B} -- but see \citealp{2003ApJ...590..348L} and \citealp{2012MNRAS.422.2246M}). 
	
	In contrast, several studies have indicated that in the most massive early-type galaxies (ETGs) the IMF slope steepens below 1M$_{\odot}$ (a `bottom-heavy' IMF). In these galaxies more dwarf stars are formed as a fraction of the total stellar population. For example, \cite{2010ApJ...721L.163A} and \cite{2012Natur.484..485C} use measurements of gravitational lensing and stellar dynamics respectively, while \cite{2011MNRAS.415..545T} uses both methods in tandem, to independently constrain the stellar mass in these systems (with the precise mass subject to an assumed dark-matter profile), finding enhanced M/L ratios in massive ETGs. An increase in the stellar M/L ratio of massive ETGs is interpreted as evidence for a non-Milky-Way-like IMF: either bottom-heavy (since the M/L of a dwarf star is much higher than that of a giant star) \textit{or} top-heavy (through a larger contribution from stellar remnants).
	
	Spectroscopic approaches to constraining the IMF provide independent support for the bottom-heavy interpretation. Such methods use measurements of absorption features in galaxy spectra which have sensitivity to stellar surface gravity (see \citealp{2012ApJ...760...71C}, hereafter CvD12b; \citealp{2012MNRAS.426.2994S}; \citealp{2012ApJ...753L..32S}; \citealp{2013MNRAS.433.3017L}; \citealp{2015MNRAS.447.1033M}; \citealp{2016ApJ...821...39M}; \citealp{2015MNRAS.452..597Z}). These features differ in strength between evolved giant stars and main-sequence dwarfs, even where the two stars have the same chemical composition and surface temperature. Their strength in the integrated light of a galaxy -- the luminosity-weighted sum of all the individual stellar spectra -- is therefore dependent on the relative number of dwarf and giant stars. These features specifically probe the contribution of dwarf stars to the total stellar light, in contrast to M/L analyses. As such, they constitute a powerful probe of the IMF in unresolved stellar populations.
	
	The main challenge for spectral synthesis methods is the presence of observational degeneracies. In particular, an increase in the abundance of a chemical element may have a similar effect to a change in the IMF on a particular feature. Breaking these degeneracies requires simultaneous consideration of a wide variety of features. While much of the previous work on gravity-sensitive features has focussed on optical spectra, there are many useful features in the near infrared that can help in disentangling the degeneracies (\citealp{2012ApJ...747...69C}, hereafter CvD12a). The development of empirical spectral libraries that cover near-infrared wavelengths, e.g. \cite{2009ApJS..185..289R}, coupled with advances in instrumentation has made study of these features increasingly feasible.
	
	Massive ETGs are thought to be assembled in two distinct stages. Gas-rich mergers of galaxies lead to rapid star formation and assembly of the cores of the ETGs at z$\sim$2--3; this starburst phase is ended by the abrupt quenching of star formation as gas is expelled from the galaxy. Slow growth via gas-poor minor mergers and much-reduced star-formation follows, through which the outer regions of the ETGs are assembled (\citealp{2010ApJ...725.2312O}; \citealp{2013MNRAS.429.2924H}). The conditions under which stars are formed in the starburst phase are very different to those encountered in the Milky-Way thin disc, which might provide a natural explanation for IMF variation \citep{2014ApJ...796...75C}. If this scenario is broadly correct the IMF will only deviate from Milky-Way-like in the cores of ETGs, leading to radial gradients in their fractional dwarf star content. Spectroscopic constraints on the IMF typically utilise spectra taken from the bright cores of ETGs. In contrast, the M/L methods typically probe the wider mass distribution. Because of this difference in scale, radial IMF gradients may provide a partial explanation for any observed mismatch between the measured M/L (from dynamics or lensing) and that inferred from a spectroscopically-derived IMF (e.g. as reported in \citealp{2014MNRAS.443L..69S}).
	
	Observations of radial variations in spectroscopic features have a long history (see, for example \citealp{1979ApJ...228..405C}, in which they were used to constrain the overall metallicity gradients of ETGs), while their use in specific attempts to constrain the dwarf star content of ETGs dates back to \cite{1986ApJ...311..637C}, in which it was concluded that ETG cores are slightly enriched in dwarf stars.
	
	Recent evidence for IMF variation in massive galaxies has given new impetus to such studies. \cite{2015MNRAS.447.1033M} present evidence from optical spectra (measurements of a variety of Mg, TiO, Na, and Ca absorption features) of three galaxies that supports a radially varying IMF in the most massive ETGs ($\sigma \sim 300$\,kms$^{-1}$). However, the small number of studies that have considered infrared spectral features (loosely defined here as wavelengths beyond the Ca\,II Triplet at 0.86$\umu$m) -- \cite{2016ApJ...821...39M} and \cite{2016MNRAS.457.1468L} -- have come to ambiguous conclusions. McConnell et al. present data from two ETGs ($\sigma$= 220\,kms$^{-1}$ and 250\,kms$^{-1}$), measuring the strengths of a variety of optical and IR absorption features, including TiO, Ca, Mg, Na, and Fe features; CN and H$\beta$; as well as the Wing-Ford FeH molecular band, which is extremely gravity sensitive (\citealp{1969PASP...81..527W}). They argue that the radial variation in these strengths is best accounted for by elemental abundance gradients, rather than requiring a varying IMF. However, they do this without quantitative fitting of models to their data and rely primarily on the lack of strong radial variation of the Wing-Ford band. 
	
	In contrast La Barbera et al. use a model-based approach, measuring a number of optical (TiO) features and the Wing-Ford band for a single extremely massive ($\sigma$= 320\,kms$^{-1}$) ETG. They find that the optical features are best fit with a radially varying IMF. However, like McConnell et al. they do not measure significant radial variation of the Wing-Ford band and invoke a more complex (broken power law) IMF functional form to account for this.
	
	We note that in neither work did the authors find the Wing-Ford band was strong in the cores of their target galaxies (compared to spectra from stellar populations formed with a Milky-Way-like IMFs), whereas in some previous work, e.g. \cite{2010Natur.468..940V}, this has been found in ETGs.
	
	In the present paper we describe first results from the KMOS Infrared Nearby Early-Type Survey (KINETyS). This is an investigation of the radial stellar population gradients in ETGs by means of IR spectroscopy, with the aim of measuring IMF trends. In this work we study eight ETGs, alleviating some of the issues with studying individual galaxies. We study a variety of gravity-sensitive features including some which were previously unexplored. We relate our measurements of these features to stellar population models in order to draw quantitative conclusions about any radial variation in the IMF and chemical abundances patterns. Our analysis emphasises the average properties of ETGs over the specific properties of individual galaxies, and is complementary to detailed studies of individual galaxies such as those described above.
	
	The paper is organised as follows: In Section 2, we first describe our sample of ETGs and the observing strategy used. We next outline our methods for reducing the data to the point where information about the stellar populations of our sample can be extracted. In Section 3 we present our measurements of the radial variations of a set of spectral features and our strategy for interpreting them. We then present the results of our analysis. In Section 4 we discuss these results and their interpretation. Finally, in Section 5 we present the conclusions of our study and note the prospects for future work.
	
	\section{Data}
	
	\subsection{Observations:}
	
	The eight nearby ($z$\,$<$\,$0.01$) ETGs observed span $\sim$150--300\,kms$^{-1}$ in velocity dispersion, with most having $\sigma \sim$ 250\,kms$^{-1}$. Seven of our sample were in the original sample of CvD12b and have a reported average central M/L of $\sim$1.6 times that expected for a Milky-Way-like IMF (we hereafter quote all M/L values according to this convention), consistent with a modestly bottom-heavy IMF (e.g. a Salpeter IMF, a power law with slope X=2.3). The most extreme reported M/L is $\sim$2, consistent with a very bottom-heavy IMF (e.g. a steeper power-law IMF with slope X=3). Full details of the sample properties are given in Table \ref{table:sample_data}. The data were gathered using the K-band Multi-Object Spectrograph (KMOS) on the VLT \citep{2013Msngr.151...21S}, primarily between 1$\mathrm{^{st}}$ April and 17$\mathrm{^{th}}$ July 2014 (Run ID: 093.B-0693(A)), with additional data for NGC$\,$1407 and NGC$\,$3379 obtained on 27$\mathrm{^{th}}$ January 2015 (Run ID: 094.B-0061(C)). 
	
\begin{figure*} 
		\centering
		\includegraphics[width=0.5\textwidth]{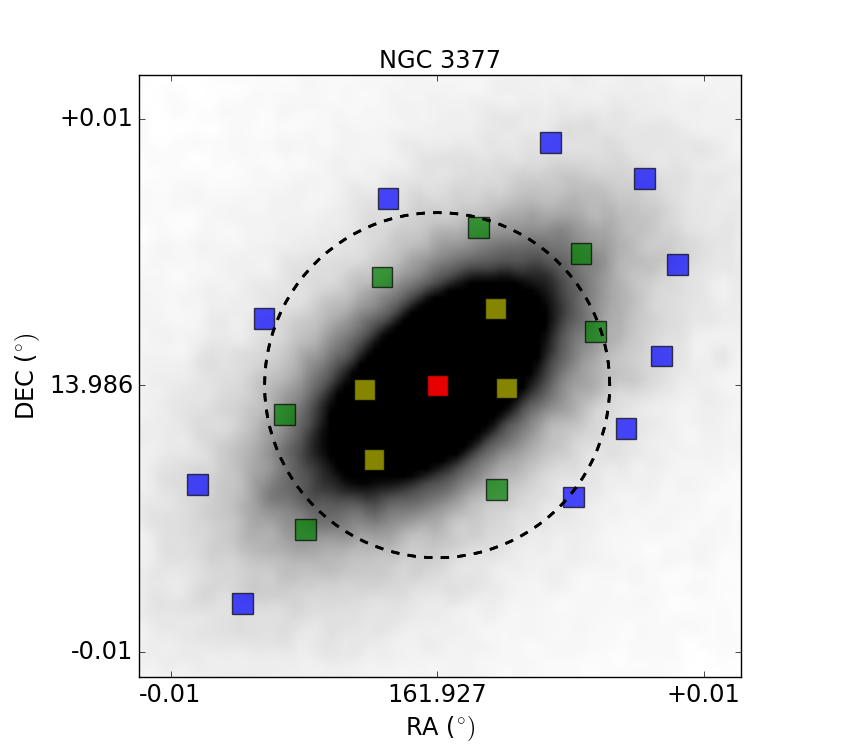}
		\includegraphics[width=0.48\textwidth]{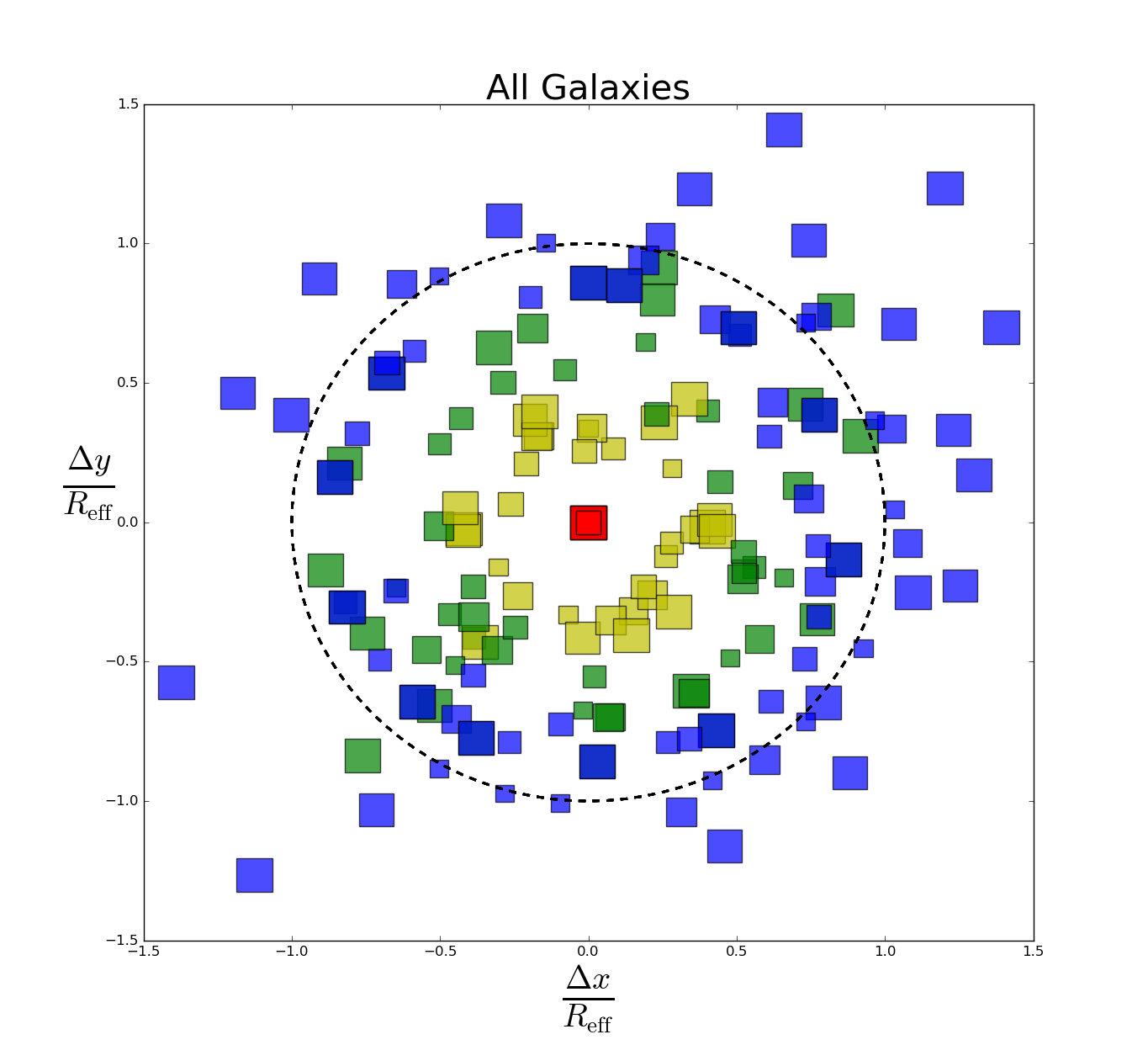}
		\caption{
			\textit{Left panel:} The KINETyS observing strategy for an example galaxy. Individual KMOS IFU fields of view (2.8$\arcsec$$\,$x$\,$2.8$\arcsec$) are shown to scale, colour coded by the isophote they are sampling. Dashed line indicates the J-band effective radius.
			\textit{Right Panel:} IFU fields for the ensemble of all galaxies scaled relative to the J-band circular effective radius. Colour scheme as in left panel.
		}
		\label{fig:observing_strategy}
	\end{figure*}

	KMOS uses 24 pick-off arms, each equipped with an integral field unit (IFU), to take spatially resolved spectra from different targets in the field of view. Each arm yields a 14$\times$14 pixel region covering a 2.8$\arcsec\times$2.8$\arcsec$ field. For these observations we deployed the pick-off arms in a sparse mosaic covering one galaxy at a time, with a single arm on the galaxy's core and the others arranged along lines of isophotal flux with major axes corresponding approximately to $\mathrm{\frac{1}{3}R_{eff}}$, $\mathrm{\frac{2}{3}R_{eff}}$, and $\mathrm{R_{eff}}$ (ellipticities are given in Table \ref{table:sample_data}). A larger number of arms were deployed to the outer regions in order to mitigate the lower surface brightness. An example of this strategy is shown in Fig. \ref{fig:observing_strategy} (left panel). The arms were allocated for observations in cross-beam switch mode, so that in each pointing of an AB pair around half were on-source. In most cases an arm was allocated to the galaxy core in both pointings.
	
	Observations were made using the IZ (0.78-1.08$\umu$m) grating and, for six of the eight galaxies, the YJ (1.03-1.34$\umu$m). Note that KMOS spectra are calibrated to vacuum wavelengths, which will be quoted throughout this work. These bands have resolving power R=3400 and R=3600 respectively. For each target we took 3$\times$450s exposures in each pointing, resulting in a total integration time of 2700s on-source. The FWHM seeing ranged from 0.8$\arcsec$ to 1.8$\arcsec$, with an average of 1.25$\arcsec$. The signal-to-noise ratio ranges from $\sim$100--270 (IZ band) and $\sim$150--370 (YJ band) in the data for the innermost region of each galaxy but falls to $\sim$5--40 in both bands at the effective radius, so those data are useful only when combined across the sample.

	\begin{table*} 
		\centering
		
		\caption{List of sample galaxies with observation details and key properties listed. Effective radii and total (J-band) magnitudes within the effective radius were extracted from 2MASS J-band images and used to calculate the mean surface brightness. Recession velocities were derived using pPXF (see Section 2.3) and $\sigma(\mathrm{R_{eff}}/8)$ values and fast/slow rotator status were taken from the ATLAS 3D survey (except for NGC\,1407 for which a value was derived from our pPXF fits). Relative M/L taken from CvD12b.}
		\label{table:sample_data}
		\begin{tabular}{clccrccccc}
			\hline
			Name     & Bands &  Seeing   & R$_{\mathrm{eff}}$ &                   cz & $\sigma(\mathrm{R_{eff}}/8)$ &   surface brightness    &  relative M/L   & ellipticity &    Notes     \\
			&       &  /arcsec  &      /arcsec       & $/\mathrm{kms^{-1}}$ &     $/\mathrm{kms^{-1}}$     & /mag$_J\,$arcsec$^{-2}$ & (Milky Way = 1) &             &  \\ \hline
			NGC$\,$0524 & IZ,YJ & 1.3,  --- &        23.3        &                 2400 &             243              &          17.8           &      1.09       & 0.00        & Fast rotator \\
			NGC$\,$1407 & IZ    & 1.4, 1.0  &        36.2        &                 1950 &             301              &          17.4           &       ---       & 0.00        & Slow rotator \\
			NGC$\,$3377 & IZ,YJ & 1.8, ---  &        23.3        &                  690 &             146              &          17.0           &      1.16       & 0.40        & Fast rotator \\
			NGC$\,$3379 & IZ    & 1.2, 0.8  &        28.5        &                  900 &             213              &          16.5           &      1.60       & 0.00        & Fast rotator \\
			NGC$\,$4486 & IZ,YJ & 1.1, 1.3  &        44.5        &                 1290 &             314              &          16.9           &      1.90       & 0.00        & Slow rotator \\
			NGC$\,$4552 & IZ,YJ & 1.1, 1.6  &        24.1        &                  390 &             262              &          16.6           &      2.04       & 0.00        & Slow rotator \\
			NGC$\,$4621 & IZ,YJ & 1.3, 1.6  &        27.7        &                  480 &             224              &          16.8           &      1.96       & 0.33        & Fast rotator \\
			NGC$\,$5813 & IZ,YJ & 1.3, 1.2  &        33.7        &                 1920 &             226              &          17.9           &      1.37       & 0.25        & Slow rotator \\ \hline
		\end{tabular}
	\end{table*} 
	
	\subsection{Data reduction}
	
	Data were initially reduced using a slightly modified version of the standard ESO data reduction pipeline, with the flat-fielding, wavelength calibration, and illumination correction steps intact, but neither the (standard-star based) telluric correction nor the sky subtraction step applied. The output data cubes were then processed to produce 1D spectra, in most cases by summing the pixel fluxes at each wavelength. In the case of the IFUs covering the galaxy core the IFU was first subdivided into pixels within 0.7$\arcsec$ of the centre and pixels further away, producing two spectra with roughly equal S/N. Since the seeing was in almost all cases better than 1.4$\arcsec$, the spreading of flux between the two spectra will not significantly affect our conclusions. Following this, a telluric correction and sky-subtraction were applied.
	
	Removal of sky emission lines was accomplished using ESO's \textsc{SkyCorr} tool (v1.0.0); see \cite{2014A&A...567A..25N}. These lines are strong in the near infrared and vary rapidly (on shorter timescales than the length of an observation) making good sky subtraction challenging. \textsc{SkyCorr} addresses this problem by rescaling physically related `families' of lines in the sky observations (in all cases taken from the same arm as the source observation but in the corresponding off-source pointing, which was pre-processed in the same way) to achieve a better match with those in the source observations prior to subtraction (see also \citealp{2007MNRAS.375.1099D} for details). We found performance was best in high S/N spectra when a first-order model for the source spectrum was first subtracted; this was accomplished by fitting a model spectrum to a conventionally sky-subtracted spectrum. Following the application of \textsc{SkyCorr} this first-order model was then added back to the spectrum. In addition, modelling and correcting for systematic variation of the sky line strengths between observations necessarily introduces additional statistical uncertainty, due to uncertainties in the parameters of the emission model. We characterised this additional uncertainty by comparing the scatter of fluxes around the mean in line regions post-subtraction with the corresponding average formal statistical uncertainty; the quadratic difference was then reincorporated into the formal uncertainty.
	
	Atmospheric absorption was corrected using ESO's \textsc{MolecFit} tool (v.1.0.2); see \cite{2015A&A...576A..77S}, \cite{2015A&A...576A..78K}. There are several wavelength ranges within our observations where this absorption is particularly strong and therefore difficult to deal with. \textsc{MolecFit} creates a best-fit atmospheric absorption profile for several key molecular absorbers, specifically O$_{2}$ and H$_{2}$O, which is then used to correct the spectrum. For each exposure the central spectrum for each galaxy (i.e. the highest S/N spectrum) was used in the fitting process and the resulting correction was then applied to the spectra taken from the other KMOS arms. In the IZ band corrections were fit to the range 8010 -- 10856\AA\, and 10210 -- 13505\AA\, in the YJ band.
	
	\textsc{MolecFit} does not introduce systematics for the majority of the absorption lines we measure since it only adjusts the spectrum in specific spectral regions. A comparison with a standard star spectrum in the spectral regions around affected features indicates differences in the applied correction of up to a few percent in any given pixel.
	
	We found that in different arms aimed at the same source relative throughput could vary significantly (10\% or more) by wavelength on scales of a few hundred pixels, in a way not corrected for by the standard pipeline. This effect, which appears to be stable over time, is most likely due to variations in the multi-layer filter coatings KMOS uses (Sharples, private communication). Using the flat lamp exposures, we were able to characterise and correct for this (multiplicative) effect by fitting a spline through the brightness profile of each arm.
	
	Additionally, a small additive effect (significant only in the low surface brightness arms) was found on similar wavelength scales, specific to individual exposures (and thus not dealt with by sky-subtraction). This is possibly due to scattered light in the instrument: the effect was present at the unilluminated edges of the detector, allowing us to characterise and subtract the variations from all arms (again, by fitting splines). The effect was not constant along the wavelength axis but appeared roughly constant across the spatial axis of the detector. 
	
	Accounting for these systematic errors ultimately leads to small corrections (less than the formal statistical uncertainty) to our measurements of spectral features (described in section 2.3), with a few exceptions where the corrections are comparable in magnitude to the formal uncertainty. Uncertainty in the application of these corrections therefore does not significantly increase the uncertainty of the measurements.
	
	\subsection{Velocity dispersions and absorption line index measurements}
	
	After applying these procedures, spectra from the multiple arms pointed at the same radial region were combined to produce a single spectrum for each physical region of each galaxy. Spectra from arms assigned to particular regions were initially summed together. Individual spectra were then divided through by this roughly combined spectrum and a sixth-order polynomial was fit to the result. Dividing each individual spectrum by the corresponding polynomial then maps each one to a common (average) continuum shape, following which they can be median-stacked in order to reduce the sensitivity of the combined spectrum to outlying pixel fluxes.
	
	Using the penalised pixel-fitting (pPXF) tool \citep{2004PASP..116..138C} we determined redshifts and velocity dispersions for these spectra (we used an approximately consistent \textit{rest-frame} wavelength fitting range of 8300--8800\AA\, and 9600-10350\AA, extending this to 10900\AA\, for those galaxies with YJ band data). For this process we used a set of template spectra based on updated versions of the models of CvD12a (hereafter ``the CvD models'') which included variations in stellar population age (3.0 to 13.5$\,$Gyr), [$\alpha$/H] (0.0 to +0.4), and the IMF (Milky-Way-like to extremely bottom-heavy).
	
	We characterised the spectra using a series of Lick-like absorption line indices, following the pseudocontinuum and feature band definitions given in CvD12a (reproduced in Table \ref{table:feature_defs}). The chosen features were selected for their strength, their IMF-sensitive nature, and their origins in a variety of different chemical species.
	
	\begin{table*}
		\centering
		\caption{List of absorption index names and definitions (vacuum wavelength definitions, given in \AA); originally from CvD12a.}
		\label{table:feature_defs}
		\begin{tabular}{l|cc|cc|cc}
			\hline
			&   Feature   &   Feature   &    Blue     &    Blue     &     Red     &     Red     \\
			Index Name                  &   (bluest   &  (reddest   &  continuum  &  continuum  &  continuum  &  continuum  \\
			& wavelength) & wavelength) &   (bluest   &  (reddest   &   (bluest   &  (reddest   \\
			&             &             & wavelength) & wavelength) & wavelength) & wavelength) \\ 
			\hline
			Na\,I (0.82\,$\umu$m)       &   8170.0    &   8177.0    &   8177.0    &   8205.0    &   8205.0    &   8215.0    \\
			Ca$\,$II (0.86$\,\upmu$m a) &   8484.0    &   8513.0    &   8474.0    &   8484.0    &   8563.0    &   8577.0    \\
			Ca$\,$II (0.86$\,\upmu$m b) &   8522.0    &   8562.0    &   8474.0    &   8484.0    &   8563.0    &   8577.0    \\
			Ca$\,$II (0.86$\,\upmu$m c) &   8642.0    &   8682.0    &   8619.0    &   8642.0    &   8700.0    &   8725.0    \\
			Mg$\,$I (0.88$\,\upmu$m)    &   8801.9    &   8816.9    &   8777.4    &   8789.4    &   8847.4    &   8857.4    \\
			FeH (0.99$\,\upmu$m)        &   9905.0    &   9935.0    &   9855.0    &   9880.0    &   9940.0    &   9970.0    \\
			Ca$\,$I (1.03$\,\upmu$m)    &    10337    &    10360    &    10300    &    10320    &    10365    &    10390    \\
			Na$\,$I (1.14$\,\upmu$m)    &    11372    &    11415    &    11340    &    11370    &    11417    &    11447    \\
			K$\,$I (1.17$\,\upmu$m a)   &    11680    &    11705    &    11667    &    11680    &    11710    &    11750    \\
			K$\,$I (1.17$\,\upmu$m b)   &    11765    &    11793    &    11710    &    11750    &    11793    &    11810    \\
			K$\,$I (1.25$\,\upmu$m)     &    12505    &    12545    &    12460    &    12495    &    12555    &    12590    \\
			Al$\,$I (1.31$\,\upmu$m)    &    13115    &    13165    &    13090    &    13113    &    13165    &    13175    \\ 
			\hline
		\end{tabular}
	\end{table*}
	
\begin{figure*}
		\centering
		\includegraphics[width=0.95\textwidth]{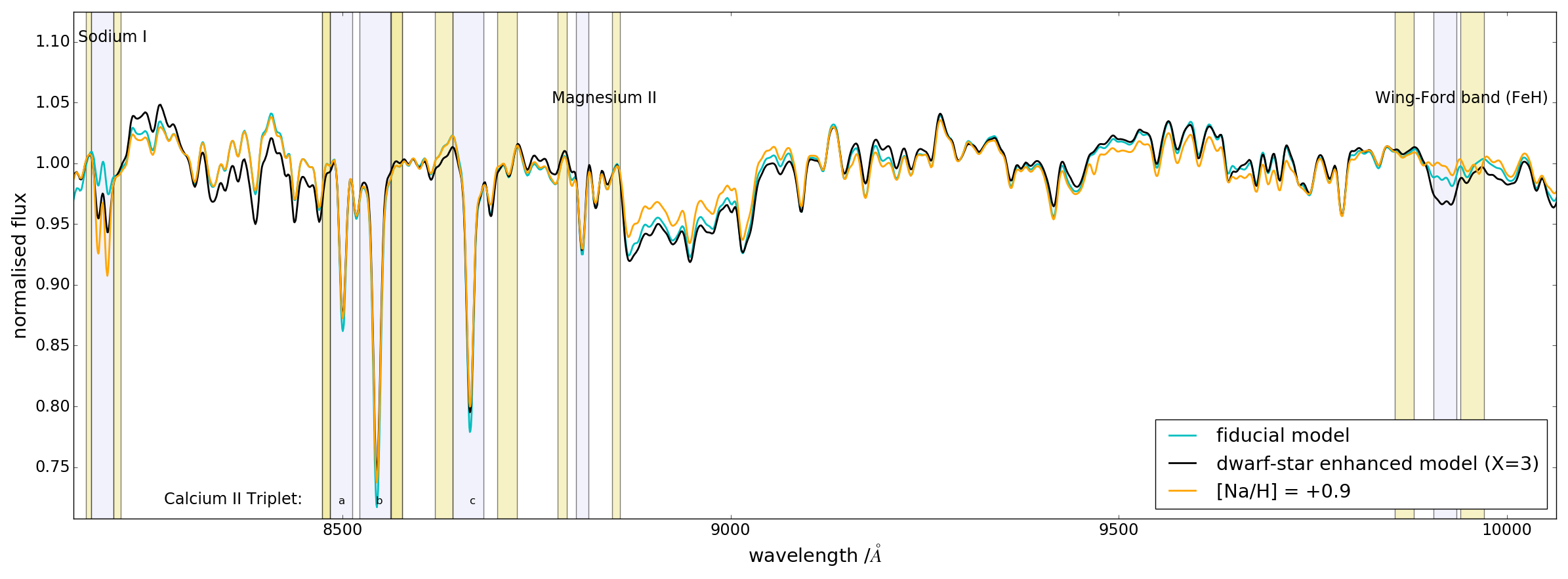}
		\includegraphics[width=0.95\textwidth]{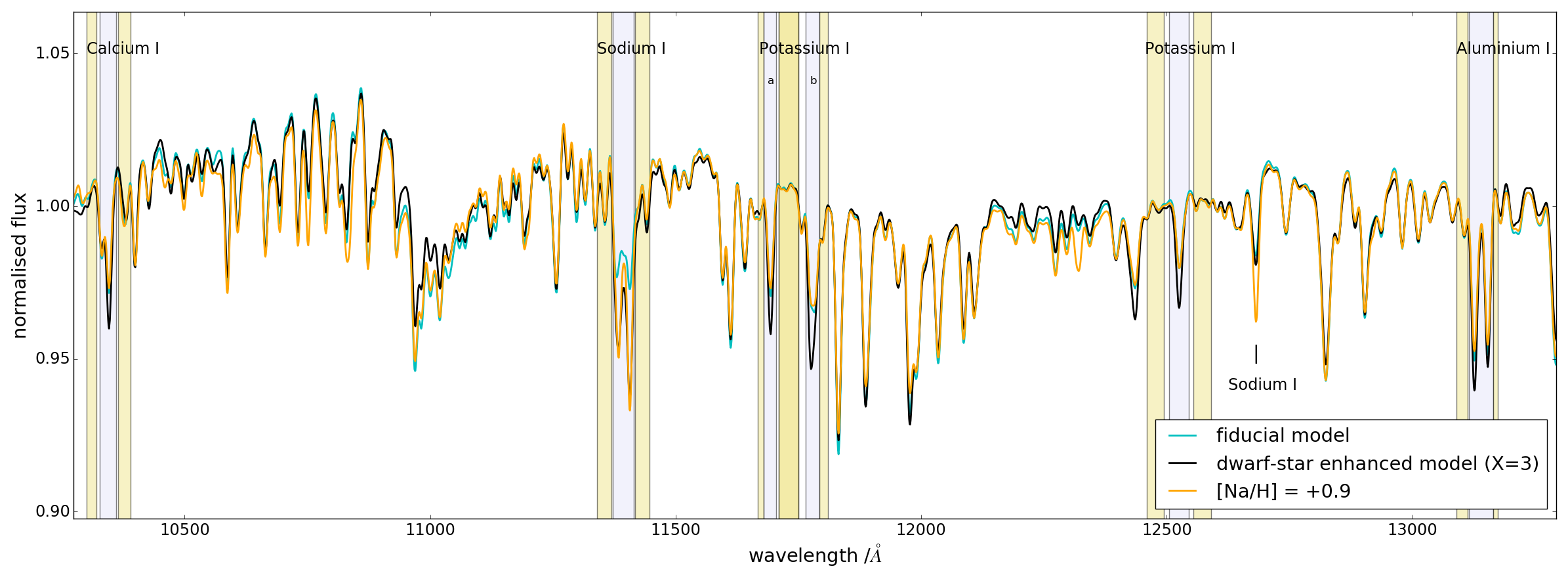}
		\caption{Three models of NIR spectra are illustrated: the solar abundance, Milky-Way-like IMF fiducial model (cyan); a model with [Na/H]$=+0.9$ (orange), and a model with a dwarf-enriched stellar population formed according to an $X=3$ IMF (black). The models are broadened to 230\,kms$^{-1}$ We include absorption index definitions for a number of absorption lines whose strength is measurably dependent on the chosen model. The equivalent width of a feature is calculated in the purple highlighted wavelength range, relative to a local (pseudo-)continuum defined in the pale side-bands and interpolated across the feature (N.B. a spline has been fit through these pseudo-continua and divided out to enable a clearer comparison of feature strengths in different models. Outside the feature definitions these models are therefore not normalised).}
		\label{fig:SSP_models}
	\end{figure*}

	Using the derived velocity dispersions and redshifts we created a suite of appropriately broadened and shifted model spectra. These were generated from the updated version of the CvD models described in \cite{2014ApJ...780...33C} and included a range of chemical abundances, population ages, and possible IMFs. In Fig. \ref{fig:SSP_models} we show three model spectra covering the same wavelength range as the data. These include a fiducial model corresponding to a single stellar population with an age of 13.5$\,$Gyr and solar chemical abundance pattern, formed according to a Milky-Way-like IMF. The other two models are variations, firstly with a large enhancement of sodium ([Na/H]$=+0.9$) and secondly with a bottom-heavy IMF (parametrized as a  power law with a slope of X=3, in comparison to X=2.35 for a Salpeter IMF). The definitions of the indices we measure are also depicted.
	
	\begin{figure*}
		\centering
		\includegraphics[width=0.97\textwidth]{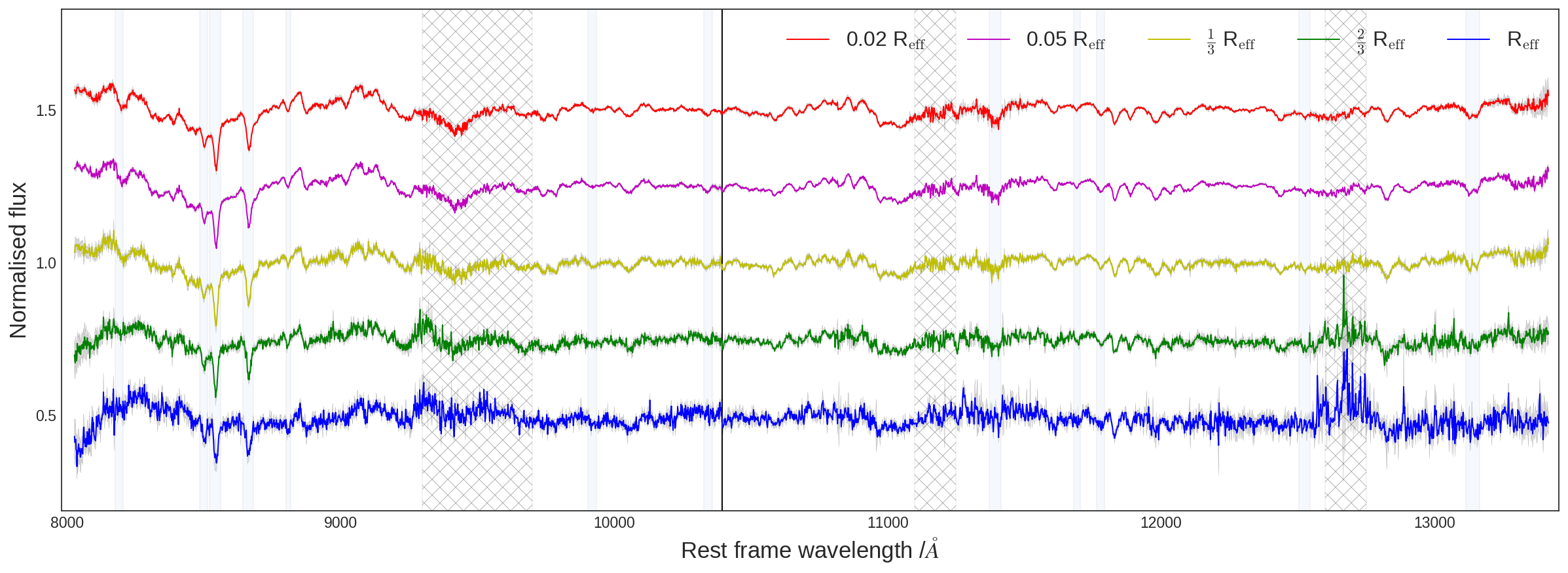}
		\caption{The data, median-stacked at fixed R/R$_{\mathrm{eff}}$, indicating the recovery of the signal in the absorption line spectra even in the outlying regions of the galaxy (feature measurement bands indicated). The colour scheme is as in Fig. 1. The hatched regions indicate particularly strong telluric absorption, visible as an increased level of noise in these regions. The dividing line between the IZ and YJ bands is also shown.}
		\label{fig:KINETYS_stacks}
	\end{figure*}
	
	Velocity broadening will change the measured equivalent width of a line due to flux being spread out of the feature definition, so it was necessary to correct the measured equivalent widths to a common velocity dispersion (230$\,$kms$^{-1}$, representative of our sample, was chosen). We measured each index in a variety of model spectra broadened to 230$\,$kms$^{-1}$, then compared these measurements to those taken from model spectra at the observed velocity dispersion. Each feature has a different intrinsic strength in each model, so these comparisons together gave us the relationship between the strength of an index at the observed velocity dispersion and its strength at 230$\,$kms$^{-1}$.
	
	We thus obtained, for each galaxy in our sample and for several locations within each galaxy, a set of measurements of absorption line strengths -- strengths which are dependent on the properties of the stellar population observed. We also derived the uncertainties on these measurements by creating a set of Monte-Carlo realisations of our spectra from the statistical uncertainties on each pixel's flux. All the measurements are reported in Appendix A.
	
	We did not correct for the relative rotational velocity of spectra from arms pointed at the same radial region during the combination procedure described at the beginning of this section. Given the use of multiple arms -- and because the intrinsic velocity dispersion of spectra from each arm is relatively large compared to the velocity shifts between them -- the resultant velocity distribution underlying the combined spectrum is indistinguishable from a Gaussian distribution in almost all cases (and therefore accurately characterised by pPXF). We tested the effect of non-Gaussianity on the most extreme case, the inner ring of four arms on NGC\,3377, but found that for synthetic spectra (with realistic noise added) broadened using a realistic kernel and the best-matching Gaussian kernel, changes to the absorption line index measurements were still well within the 1-$\sigma$ measurement uncertainties even in this extreme case. The realistic kernel we used was estimated using velocities and velocity dispersions from the ATLAS$^{3D}$ survey (\citealp{2004MNRAS.352..721E}, \citealp{2011MNRAS.413..813C}). Unfortunately velocity maps are not available out to the effective radius for all galaxies in our sample and S/N is not high enough to fit kinematics to spectra from individual arms. This prevents us from removing relative rotational velocities prior to combination, as would otherwise be optimal.
	
	Before interpreting the measured index equivalent widths we assessed the impact of contamination of our absorption features by active galactic nuclei (AGN). Four of our sample are known to contain low-ionisation nuclear emission regions (LINERs): NGC$\,$3379, NGC$\,$4486, NGC$\,$4552, and NGC$\,$5813 (see for example \citealp{1997ApJS..112..315H}). However, we found that contamination was significant in only one galaxy (NGC\,4486) and this could be straightforwardly corrected for most absorption features. Details of this analysis and correction can be found in Appendix B. The small number of feature measurements we were unable to correct for this galaxy are not included in the subsequent analysis.
	
	\section{Interpreting absorption index strengths}
	
	As described in the Introduction, some measured absorption features are sensitive to the IMF. However, the effect of dwarf enrichment of the stellar population on any given feature is always degenerate with other effects (e.g. chemical abundances). This not only follows from the individual dependence of particular atomic/molecular transitions on multiple parameters, but also from blending with neighbouring features (due to velocity broadening).
	
	In Fig. \ref{fig:KINETYS_stacks} we show spectra from each of the radial extraction regions, median-stacked over the eight galaxies (after normalising the continua) in order to suppress noise. The features that were highlighted in Fig. \ref{fig:SSP_models} are readily apparent.
	
	We created these stacked spectra in the following way: first, we took the spectra from the innermost extraction region (the central 0.7\arcsec\, of the subdivided central IFU) for each galaxy, shifted them into the rest-frame and binned them onto a common wavelength grid. We then divided out their relative continuum variation (by first fitting a 6$^{th}$ order polynomial to the ratio of each spectrum with the mean spectrum and dividing through by this polynomial) and then evaluated the median flux at each wavelength. Errors were then created by bootstrap resampling of the input spectra, since galaxy-galaxy variations are larger than the statistical uncertainty on any given spectrum. We followed a similar procedure for the other spectra, but in each case divided the spectra by that taken from the innermost extraction region prior to stacking (i.e. we stacked the variations in the spectra). These `variation-stacks' were then multiplied by the previously-calculated central stacked-spectrum. This procedure is intended to maximise the radial variation signal by dividing out any constant offset in the strength of spectral features between different galaxies in the stack.
	
	The main goal of this paper is to constrain the radial variation of the stellar populations within a sample of several large, nearby ETGs. This is accomplished by modelling the behaviour of a subset of the spectral features we measure. 
	
	We analyse our data in two complementary ways. First, we consider, for a subset of our sample of galaxies, the strength of various absorption features in the three inner radial bins. We aim to use these to characterise the stellar populations of these galaxies, thereby inferring how the properties of their stellar populations vary with radius. Secondly, we consider the variation of spectral features between the stacked spectra created by the procedure described above. By stacking in the rest frame we gain by suppressing systematic effects and are able to quantify any radial gradients by using the full range of spectra available to us, at the cost of `washing out' information about the individual galaxies. We therefore treat the stacked spectra as being representative of an average massive ETG at a given fraction of the effective radius. We measure indices for these stacked spectra in precisely the same way as for the individual spectra, including the correction of all measurements to a common velocity dispersion (the best fit velocity dispersion at each radius is given in Table \ref{table:stack_disps} for both wavebands) and the inference of errors through multiple Monte-Carlo realisations of the data. These measurements are recorded in Appendix A alongside those for individual galaxies.
	
	\begin{table}
		\centering
		\caption{ Velocity dispersion of the stacks (corrected for instrumental broadening) }
		\label{table:stack_disps}
		\begin{tabular}{lcc}
			\hline
			Extraction radius R/R$_{\rm eff}$ & $\sigma_{\rm IZ}$ /kms$^{-1}$ & $\sigma_{\rm YJ}$ /kms$^{-1}$ \\ \hline
			0.02                              &         219 $\pm$ 19          &         235 $\pm$ 20          \\
			0.05                              &          217 $\pm$ 6          &          197 $\pm$ 8          \\
			0.33                              &         192 $\pm$ 12          &          167 $\pm$ 7          \\
			0.67                              &         175 $\pm$ 13          &         172 $\pm$ 18          \\
			1.00                              &         146 $\pm$ 30          &         180 $\pm$ 15          \\ \hline
		\end{tabular}
	\end{table}
	
	\subsection{Radial variations in the measured index strengths}
	
\begin{figure*}
		\centering
		\includegraphics[width=0.99\textwidth]{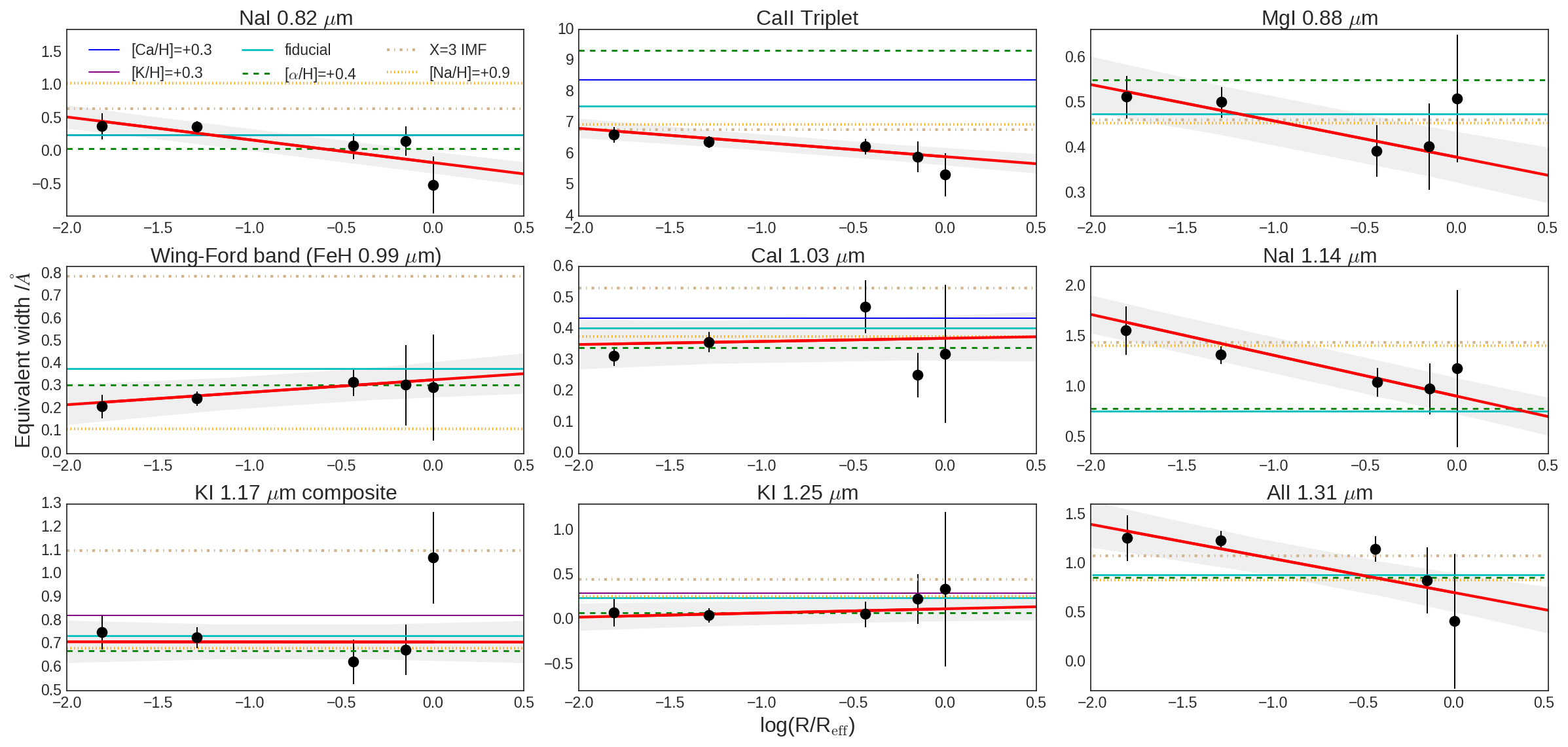}
		\caption{The index trends with log(R/R$_{\mathrm{eff}}$). The index measurements for each radial extraction region (corrected to 230\,kms$^{-1}$ in all cases) are shown in black with errors. The best fit trends are in red with the calculated 1$\sigma$ uncertainty region shown (grey shaded). The SSP model predictions are denoted by horizontal lines (Cyan: fiducial model; Green: [$\alpha$/H]=+0.4; Tan: X=3 IMF; Orange: [Na/H]=+0.9; Dark blue: [Ca/H]=+0.3 (but note that Ca is also an $\alpha$-element so scales with [$\alpha$/H] too);  Purple: [K/H]=+0.3)}
		\label{fig:stacktrends}
	\end{figure*}

	We first consider line strength trends with log(R/R$_{\mathrm{eff}}$) for the stacked spectra. It is not clear \textit{a priori} whether parametrizing the radial coordinate linearly or logarithmically is more `natural', but we note that the logarithmic description gives more leverage over the fit to the high S/N data from the central IFU. The trends are shown in Fig. \ref{fig:stacktrends}, in which the measurements from the stacked spectra are displayed along with the best fitting trend line. The error bars were created by bootstrap resampling as detailed above, so indicate the spread of measurements across the sample of galaxies from which the stacks were created. Also shown (as horizontal lines) are the predictions of the CvD models for comparison. In this section we report these empirical correlations, setting aside explicit model fitting until Section 3.2.
	
	\begin{table}
		\centering
		\caption{Measured absorption indices and their best-fit radial gradients. The most significant trends are highlighted in bold.}
		\label{table:empirical_trends}
		\begin{tabular}{lc}
			\hline
			Index Name       &     best-fit gradient /\AA     \\
			& (log(R/R$_{\mathrm{eff}}$)) \\ \hline
			Na$\,$I 0.82$\umu$m   & \textbf{--0.34 $\pm$ 0.05} \\
			Ca$\,$II Triplet    & \textbf{--0.45 $\pm$ 0.14} \\
			Mg$\,$I 0.88$\umu$m   & \textbf{--0.08 $\pm$ 0.02} \\
			Wing-Ford band     &      +0.06 $\pm$ 0.04      \\
			Ca$\,$I 1.03$\umu$m   &     +0.01  $\pm$ 0.03      \\
			Na$\,$I 1.14$\umu$m   & \textbf{--0.40 $\pm$ 0.05} \\
			K$\,$I 1.17$\umu$m a+b &     --0.00  $\pm$ 0.04     \\
			K$\,$I 1.25$\umu$m   &      +0.05 $\pm$ 0.06      \\
			Al$\,$I 1.31$\umu$m   &     --0.34 $\pm$ 0.12      \\ \hline
			&
		\end{tabular}
	\end{table} 
	
	In Table \ref{table:empirical_trends} we provide the best-fitting linear trends of index strength with log(R/R$_{\mathrm{eff}}$). Those with particular statistical significance ($>3\sigma$) are highlighted in bold. We measure negative radial trends in the Ca$\,$II Triplet and the Na$\,$I 1.14$\umu$m line (i.e. these lines are significantly weaker at the effective radius than they are in the core). More marginal trends are observed in some other indices, but we note in particular that our measurements of the IMF-sensitive Wing-Ford band are consistent with no radial trend.
	
	The uncertainties listed in Table \ref{table:empirical_trends} account for the uncertainties on each pixel in the stacked spectra and the (small) uncertainties introduced by the velocity dispersion correction to the stacked spectra, as well as accounting for the uncertainty in the best-fit gradient due to the scattering of the points around a straight-line relationship.

	\subsection{Model fitting to the measured index strengths}
	
	The strong radial variations in certain measured spectral features appear to indicate significant radial variation in the stellar populations of our sample of ETGs. We now describe our method for fitting models to the data.
	
	Model parameter estimation is accomplished in this work using the \texttt{emcee} routine \citep{2013PASP..125..306F}, a Markov-Chain Monte Carlo (MCMC) ensemble sampler (based on the method described in \citealp{Goodman_and_Weare}) used to characterise the posterior probability distribution for a chosen model, given data and statistical uncertainties. 
	
	The (log-)likelihood function for our data, $d$, given a choice of model parameters $\theta$ and uncertainties $\sigma$ is
	
	$$
	\mathrm{ln}\, p(d|\theta , \sigma) = -\frac{1}{2}\sum_i \left(\frac{d_i - m_i(\theta)}{\sigma_i}\right)^2
	$$
	
	Given a prior probability distribution, $p(\theta)$, we can use this to evaluate the posterior probability function $p(\theta|d,\sigma)$. In this work we use a set of `top-hat' constraints for $p(\theta)$ which are flat (i.e. uninformative) within some plausible range before cutting off sharply. The method allows us to efficiently find the choice of model parameters $\theta$ required to maximize the posterior probability function and determine how well constrained this solution is.
	
	Using this method requires a number of choices to be made other than what probabilistic prior to use, namely which spectral features to include in the fits and which stellar population parameters to fit for (or more generally, what model to use). For the former, we use a set of features chosen for their strength and their sensitivity to the IMF and key chemical abundances. The sensitivities of these features in the CvD models then suggest a choice of parameters we ought to be able to constrain.
	
	\subsubsection*{Spectral features used to constrain the model}
	
	In order to break the degeneracies between different stellar population parameters more effectively, we make use of some archival optical line strength maps for the Mgb, H$\beta$, and Fe\,5015 indices. These were made available by \cite{2015MNRAS.448.3484M} as part of the ATLAS$^{3D}$ survey (which includes all of our sample, except NGC\,1407). The optical features collectively act as powerful constraints on [$\alpha$/H], [Fe/H], and stellar population age, all of which are key parameters in the CvD models. Using the 2D line-strength maps allows us to carefully choose appropriate radial extraction regions to match our own data. 
	
	In addition to the optical measurements derived for individual galaxies, we estimated complementary measurements for the stacked spectra. For the central radial extraction region we used the median optical measurement, inflating the measurement uncertainty to the uncertainty of the next most central radial bin to compensate for any aperture mismatch. In the other radial extraction regions we measured the ratio of the line strengths with the central line strengths and applied the median \textit{ratio} with the central estimate to create estimates for the other radii (in an analogous way to our procedure for creating the stacked spectra).
	
	Our wavelength range contains two sodium absorption features, Na$\,$I 0.82$\umu$m and Na$\,$I 1.14$\umu$m. Both are primarily sensitive to the IMF and to the abundance of sodium \citep{2012ApJ...747...69C}. 
	
	The 0.82$\umu$m feature has made a significant contribution to measurements of bottom-heavy IMFs, dating back to \cite{1971ApJS...22..445S}. The 1.14$\umu$m sodium feature is relatively unexplored. \cite{2015MNRAS.454L..71S} reported that in ETGs it is often much stronger than would be expected from the fiducial SSP model. In the CvD models the 1.14$\umu$m feature has particularly high sensitivity to the IMF in comparison with its sensitivity to sodium abundance.
	
	The Ca\,II 0.86$\umu$m Triplet is the strongest absorption feature we measure. The feature is weakened in models with bottom-heavy IMFs and/or enhanced [Na/H] and is of course sensitive to [Ca/H]. Calcium is formally an $\alpha$ element, but its abundance does not always track [$\alpha$/H] in $\alpha$-enhanced stellar populations (e.g. \citealp{2003MNRAS.343..279T}).
	
	We measure a second Ca feature, a weak line at 1.03$\umu$m. Unlike the other features mentioned so far, it is insensitive to the sodium abundance -- see \citealp{2012MNRAS.426.2994S}. This feature is complementary to the Ca\,II Triplet in the sense that variation in the IMF slope will alter their strengths in opposite ways (the triplet is weakened in the CvD models in the bottom-heavy case, while the Ca$\,$I feature gets stronger). In contrast both are strengthened if [Ca/H] is enhanced. Thus, taken in concert, the two features in principle provide a powerful constraint on IMF variations.
	
	The Wing-Ford band is a molecular absorption band at 0.99$\umu$m associated with the FeH molecule. It is particularly sensitive to the IMF, being present in cool dwarf stars but not giants \citep{1977A&A....56....1N} and also weakens as [Na/H] increases.
	
	We have measured the K\,I 1.17$\umu$m composite feature and the K\,I 1.25$\umu$m line, neither of which have previously been studied in detail. Both are primarily sensitive to the IMF and to [K/H], while the K\,I 1.25$\umu$m line is dramatically weakened as [$\alpha$/H] increases (due to changes in the local continuum behaviour).
	
	We measure the Al\,I 1.31$\umu$m doublet, which is sensitive to the IMF but is not well explored. Because the CvD models do not explictly account for variations in [Al/H] at the time of writing, we do not attempt to draw quantitative conclusions from this feature, but include our measurements for completeness.
	
	Taken together, these features have significant sensitivity to the abundance of $\alpha$-elements (in particular Ca), Fe, Na, Ca, K, and the IMF.

	\subsubsection*{Choosing an appropriate model}
	
	Our aim is to find a model which reproduces the measured equivalent widths of the spectral features just described. The CvD models can be used to evaluate the model equivalent widths for a variety of different stellar populations with different ages and chemical abundances, formed according to various IMFs. We seek a model function to predict equivalent widths given a certain set of parameters which encapsulate these possible variations in the stellar populations we observe. To do this we interpolate the model grid of predicted equivalent widths using a set of splines; this works best when the grid is well-behaved (i.e. the predicted equivalent widths do not change sharply with respect to the chosen model parameters). Thus, we specified the model parameters that describe the stellar populations in such a way as to roughly linearise the variation in predicted equivalent widths with respect to these (without treating these variations as precisely linear in all cases).
	
	For the IMF, the parameter we choose to linearise the model behaviour is $f_{\mathrm{dwarf}}$, the fraction of the total continuum (J-band) \textit{luminosity} contributed by stars with masses below M$_{\mathrm{dwarf}}$ = 0.5\,M$_{\odot}$ (different choices for this definition, e.g. 0.2\,M$_{\odot}$, make a negligible difference to our analysis). Throughout this work we express f$_{\rm dwarf}$ as a percentage of the total light. We compute this for the Chabrier IMF and for a variety of IMFs with a single-slope below 1\,M$_{\odot}$ via the BaSTI \citep{2013A&A...558A..46P} isochrones, assuming an old (age of 13.5\,Gyr) stellar population. As detailed in Appendix C, we find an approximately linear relationship between this quantity and the measured strengths of all our absorption features in the SSP models. Using $f_{\mathrm{dwarf}}$ allows us to treat models which assume different IMF functional forms equally by capturing their differences in a single linear parameter. For example, the difference in $f_{\mathrm{dwarf}}$ between a Chabrier IMF and an X=3 power-law IMF is 14.7\% under our definition of $f_{\mathrm{dwarf}}$. We note that $f_{\mathrm{dwarf}}$ is not completely independent of IMF shape -- the linear relationships we find can be broken by sufficiently exotic IMF variations and this should be borne in mind when interpreting our results.
	
	Parametrizing the elemental abundance variations of the stellar populations is done using the relative logarithmic abundances (e.g. [$\alpha$/H]). While the behaviour of some indices is not perfectly linear in these quantities, it is usually sufficiently close that the effects of nonlinearity do not substantially affect our conclusions.
	
	We make the standard assumption (e.g. CvD12b) that the behaviours of the parameters in our model are not strongly coupled, i.e. that variation of one parameter does not markedly alter the spectroscopic response of any spectral feature with respect to another parameter.
	
	In summary, our model is parametrized principally by [$\alpha$/H], [Fe/H], [Na/H], [Ca/H], [K,H], and f$_{\rm dwarf}$ (a parameter that expresses variations in the low-mass end of the IMF). Given that Ca is already accounted for in [$\alpha$/H], fitting for [Ca/H] in fact reduces to fitting for [Ca/$\alpha$]. Details of the priors imposed on these parameters are given in Table \ref{table:parsandpriors}.

	\begin{table}
		\centering
		\caption{ Model parameters and upper and lower cutoffs for the imposed priors. }
		\label{table:parsandpriors}
		\begin{tabular}{lcc}
			\hline
			Model Parameter      & Minimum value & Maximum value \\ \hline
			$\rm [\alpha/H]$     &     --0.4     &      0.7      \\
			$\rm [Fe/H] $        &     --0.5     &      0.4      \\
			f$_{\mathrm{dwarf}}$ &     \gap{}4.3\%     &    \gap{}30.0\%     \\
			$\rm [Na/H] $        &     --1.3     &      1.3      \\
			$\rm [Ca/\alpha] $   &     --0.45     &      0.45      \\
			$\rm [K/H] $         &     --1.3     &      1.3      \\ \hline
		\end{tabular}
	\end{table}
	
	\subsection{Results}
	
	We now present the results of our analysis both for five individual galaxies (NGC\,0524, NGC\,3377, NGC\,4552, NGC\,4621, and NGC\,5813) and the stacked spectra described in Section 3. These individual galaxies have a full set of (non-AGN contaminated) optical, IZ, and YJ data and therefore we can place strong constraints [Fe/H] and [$\alpha$/H], which is essential to the proper inference of the IMF (we did not fit the individual spectra of the other galaxies because they lacked these constraints).
	
	Each MCMC run consisted of 100 `walkers' exploring the posterior probability distribution over 5000 steps (i.e. a total of 500,000 samples), with an initial 200 of these steps (20,000 samples) comprising the `burn-in' stage discarded in the usual way. The initial positions of the walkers were drawn at random from the prior probability distribution and we examined the convergence of the chains to ensure the equilibrium distribution had been reached.
	
\begin{figure*}
		\centering
		\includegraphics[width=0.98\textwidth]{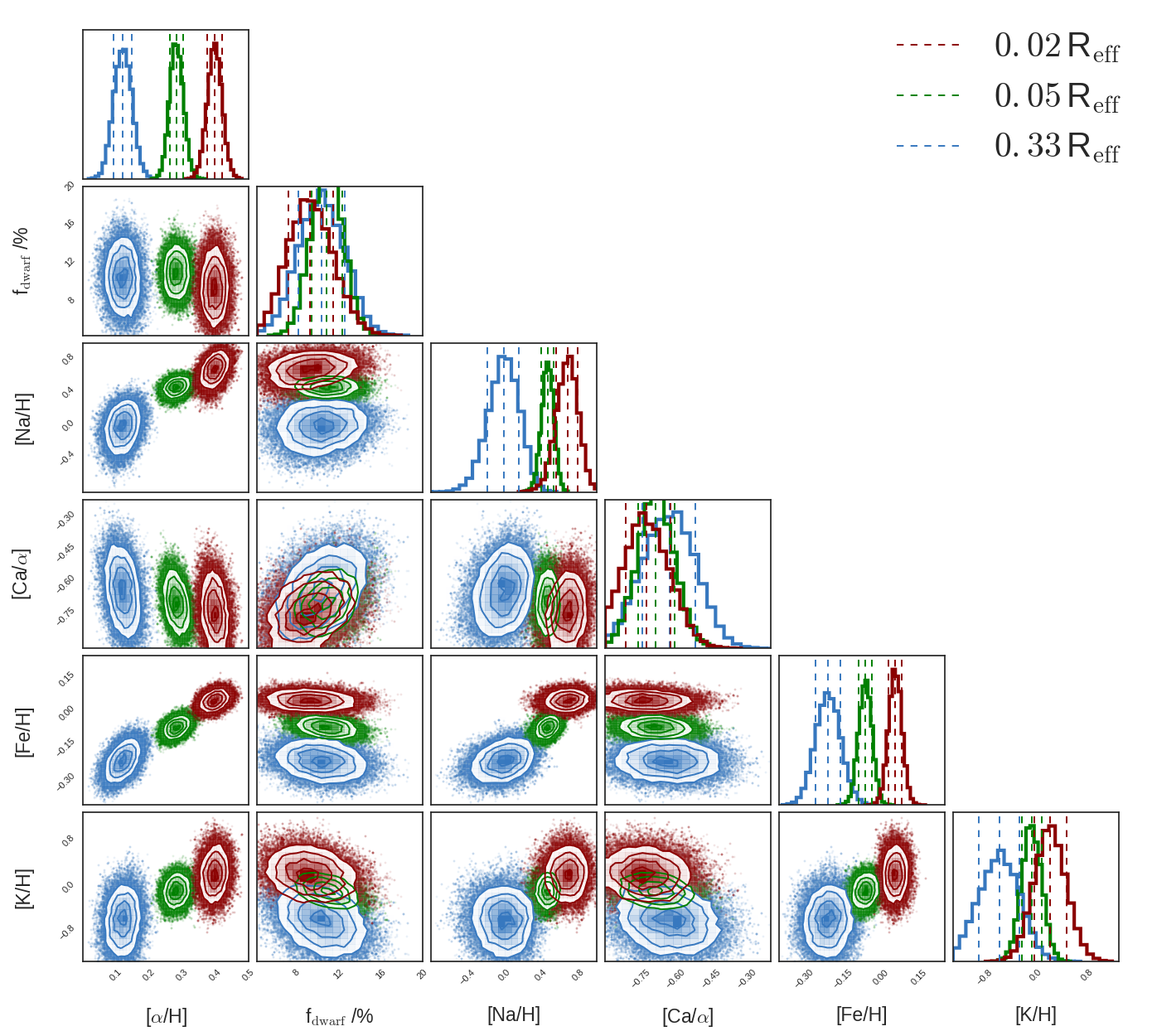}
		\caption{MCMC output for the stacked spectra at three different radii (the two outermost stacks have been omitted for clarity). The 1D distributions are plotted along the diagonal, with the vertical dashed lines showing the 16$^{\rm th}$, 50$^{\rm th}$, and 84$^{\rm th}$ percentiles. The 2D (covariance) distributions are also plotted, with contours at $\frac{1}{2}\sigma$ intervals. The radial changes in chemical abundance are clearly visible and well constrained in most cases. The [Ca/$\alpha$] ratio appears to be constant and nor do we see a radial gradient in the IMF (parametrized by f$_{\rm dwarf}$).}
		\label{fig:MCMC_grad}
	\end{figure*}

	In Figure \ref{fig:MCMC_grad} we display some example output from our method, indicating the parameter covariances derived from our method for three stacked spectra (innermost three extraction radii).
	
\begin{figure*}
		\centering
		\includegraphics[width=0.98\textwidth]{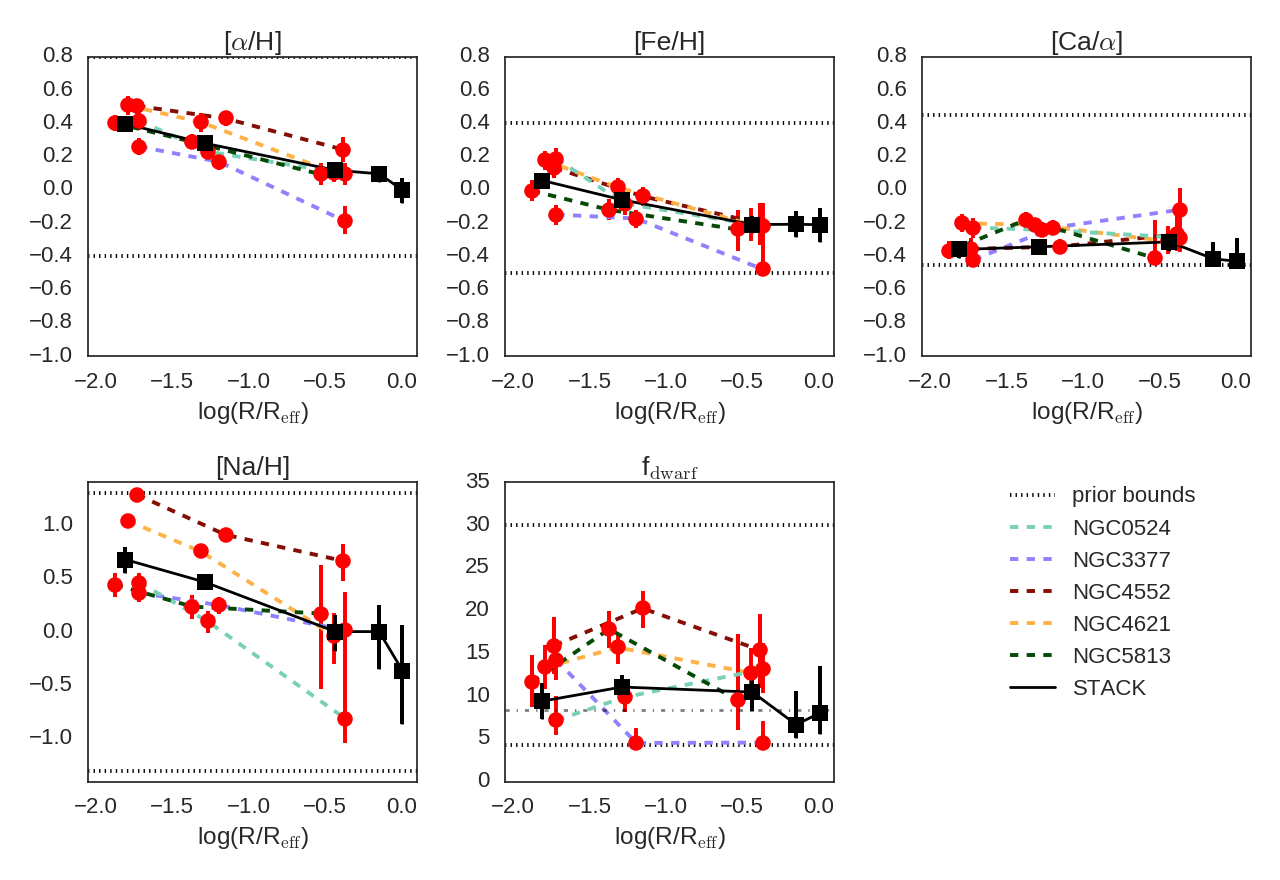}
		\caption{ Results of parameter estimation for five galaxies (red points) with IZ and YJ band data and ATLAS$^{3D}$ optical data at three different radial extraction regions. Also shown are the results derived from the stacked spectra (black squares) at five different radii. The log-radial chemical abundance gradients appear rather uniform, while there is some diversity for f$_{\rm dwarf}$. The dotted horizontal lines indicate the prior constraints on the model parameters, while the dash-dotted line on the f$_{\rm dwarf}$ panel displays the particular case, for reference, of a constant IMF with the Salpeter slope. }
		\label{fig:individual_grads}
	\end{figure*}

	Our results are shown in Figure \ref{fig:individual_grads}. We are able to demonstrate the existence of chemical abundance gradients and, moreover, show that these gradients are rather uniform between galaxies, even though those galaxies have disparate central chemical abundance patterns. The stacked spectra appear to represent this behaviour well: the parameters inferred appear to be close to the average in the inner three radial bins and the gradients in the stack appear similar to those in individual galaxies. We infer a strong metallicity gradient with [$\alpha$/H] falling by $\sim$0.2 per decade in R/R$_{\rm eff}$. [$\alpha$/Fe] appears to be constant with radius ($\sim+0.35$) and Ca appears to track Fe rather than the other $\alpha$-elements, with [Ca/$\alpha$] roughly constant at --0.35. Meanwhile, [Na/H] exhibits an extremely strong radial gradient, the steepness of which is remarkably consistent between galaxies. 
	
	By contrast, the radial behaviour of the IMF which we infer appears to be complex. From the stacked spectra, we infer an IMF which is moderately bottom heavy (corresponding to a Salpeter IMF) and roughly constant with radius. The individual galaxies however exhibit a wide variety of radial IMF trends; some we infer to host extremely bottom-heavy IMFs while others are more consistent with Salpeter in the core. In some galaxies there is evidence of radial variation, but this is by no means true in all cases.
	
	Our full results for individual galaxies are given in Tables \ref{table:parsandresults1}, \ref{table:parsandresults2}, and \ref{table:parsandresults3}. Results for the stacked spectra are given in Table \ref{table:stackparsandresults}. For the stacked spectra, we also derive the best-fitting parameter gradients, e.g. $ \rm \Delta [\alpha/H] / \Delta log(R/R_{eff}) $; these are also given in Table \ref{table:stackparsandresults}.
	
	\begin{table*}
		\centering
		\caption{ Model parameters and best fit values (R $< 0.02$\,R$_{\rm eff}$)}
		\label{table:parsandresults1}
		\begin{tabular}{lcccccc}
			\hline
			Galaxy    &   $\rm [\alpha/H]$   &       $\rm [Fe/H] $        &     f$_{\mathrm{dwarf}}$     &    $\rm [Na/H] $     &   $\rm [Ca/\alpha] $   & reduced $\chi^2$ \\ \hline
			NGC\,0524 & 0.42\err{0.05}{0.05} & \gap{}0.19\err{0.06}{0.07} & \phantom{0}7.2\err{2.9}{1.8} & 0.46\err{0.09}{0.11} & --0.23\err{0.06}{0.06} &  3.22 \\
			NGC\,3377 & 0.26\err{0.05}{0.05} &   --0.15\err{0.06}{0.06}   &      14.2\err{2.4}{2.3}      & 0.37\err{0.09}{0.09} & --0.42\err{0.05}{0.02} &  6.21 \\
			NGC\,4552 & 0.50\err{0.04}{0.05} & \gap{}0.13\err{0.05}{0.06} &      15.9\err{3.3}{3.4}      & 1.28\err{0.01}{0.07} & --0.36\err{0.07}{0.05} &  8.03 \\
			NGC\,4621 & 0.51\err{0.06}{0.06} & \gap{}0.18\err{0.06}{0.06} &      13.4\err{2.6}{2.6}      & 1.04\err{0.04}{0.05} & --0.20\err{0.06}{0.05} &  9.34 \\
			NGC\,5813 & 0.40\err{0.05}{0.05} &   --0.01\err{0.06}{0.06}   &      11.7\err{3.2}{2.9}      & 0.44\err{0.10}{0.11} & --0.37\err{0.05}{0.05} &  3.82 \\ \hline
		\end{tabular}
	\end{table*}
	
	\begin{table*}
		\centering
		\caption{ Model parameters and best fit values (R = 0.02--0.05\,R$_{\rm eff}$) }
		\label{table:parsandresults2}
		\begin{tabular}{lcccccc}
			\hline
			Galaxy    &   $\rm [\alpha/H]$   &       $\rm [Fe/H] $        &     f$_{\mathrm{dwarf}}$     &    $\rm [Na/H] $     &   $\rm [Ca/\alpha] $   & reduced $\chi^2$\\ \hline
			NGC\,0524 & 0.23\err{0.04}{0.04} &   --0.08\err{0.06}{0.07}   & \phantom{0}9.9\err{1.7}{1.7} & 0.10\err{0.09}{0.11} & --0.24\err{0.05}{0.04} & \phantom{0}2.93 \\
			NGC\,3377 & 0.17\err{0.06}{0.05} &   --0.17\err{0.05}{0.06}   & \phantom{0}4.5\err{1.7}{0.2} & 0.25\err{0.08}{0.09} & --0.23\err{0.05}{0.05} & 16.30 \\
			NGC\,4552 & 0.43\err{0.04}{0.04} &   --0.04\err{0.06}{0.06}   &      20.3\err{2.1}{2.3}      & 0.91\err{0.05}{0.05} & --0.34\err{0.04}{0.04} & \phantom{0}8.54 \\
			NGC\,4621 & 0.41\err{0.05}{0.06} & \gap{}0.01\err{0.05}{0.06} &      15.7\err{1.7}{1.9}      & 0.75\err{0.04}{0.06} & --0.21\err{0.04}{0.05} & 11.2 \\
			NGC\,5813 & 0.29\err{0.05}{0.04} &   --0.12\err{0.06}{0.06}   &      17.8\err{2.1}{2.2}      & 0.24\err{0.11}{0.11} & --0.18\err{0.04}{0.05} & \phantom{0}1.29 \\ \hline
		\end{tabular}
	\end{table*}
	
	\begin{table*}
		\centering
		\caption{ Model parameters and best fit values (R = $\frac{1}{3}$\,R$_{\rm eff}$) }
		\label{table:parsandresults3}
		\begin{tabular}{lcccccc}
			\hline
			Galaxy    &      $\rm [\alpha/H]$      &     $\rm [Fe/H] $      &     f$_{\mathrm{dwarf}}$     &       $\rm [Na/H] $        &   $\rm [Ca/\alpha] $   & reduced $\chi^2$\\ \hline
			NGC\,0524 & \gap{}0.10\err{0.06}{0.07} & --0.22\err{0.14}{0.12} &      13.1\err{3.0}{2.8}      &   --0.81\err{0.34}{0.23}   & --0.29\err{0.11}{0.09} & 3.98 \\
			NGC\,3377 &   --0.19\err{0.09}{0.08}   & --0.48\err{0.14}{0.02} & \phantom{0}4.6\err{2.5}{0.2} & \gap{}0.01\err{0.36}{0.49} & --0.12\err{0.13}{0.15} & 1.18 \\
			NGC\,4552 & \gap{}0.24\err{0.07}{0.07} & --0.20\err{0.12}{0.13} &      15.4\err{4.3}{4.4}      & \gap{}0.66\err{0.16}{0.19} & --0.27\err{0.12}{0.10} & 1.31 \\
			NGC\,4621 & \gap{}0.1\err{0.06}{0.06}  &  --0.21\err{0.1}{0.1}  &      12.7\err{3.0}{3.0}      &   --0.04\err{0.21}{0.27}   & --0.31\err{0.09}{0.08} & 3.36 \\
			NGC\,5813 & \gap{}0.09\err{0.07}{0.06} & --0.24\err{0.12}{0.13} & \phantom{0}9.5\err{7.7}{3.4} & \gap{}0.17\err{0.45}{0.71} & --0.41\err{0.23}{0.03} & 0.32 \\ \hline
		\end{tabular}
	\end{table*}

	\begin{table*}
		\centering
		\caption{ Model parameters and best fit values for stacked spectra at five radii }
		\label{table:stackparsandresults}
		\begin{tabular}{lcccccc}
			\hline
			Extraction Region                &   $\rm [\alpha/H]$   &       $\rm [Fe/H] $        &     f$_{\mathrm{dwarf}}$     &       $\rm [Na/H] $        &   $\rm [Ca/\alpha] $   & reduced $\chi^2$ \\ \hline
			R $< 0.02$\,R$_{\rm eff}$        & 0.40\err{0.02}{0.02} & \gap{}0.06\err{0.03}{0.03} & \phantom{0}9.4\err{2.1}{2.1} & \gap{}0.68\err{0.12}{0.12} & --0.36\err{0.05}{0.05} & 0.84 \\
			R = 0.02--0.05\,R$_{\rm eff}$    & 0.28\err{0.02}{0.02} &   --0.06\err{0.03}{0.03}   &      11.1\err{1.5}{1.5}      & \gap{}0.47\err{0.07}{0.07} & --0.34\err{0.04}{0.04} & 2.00 \\
			R = $\frac{1}{3}$\,R$_{\rm eff}$ & 0.12\err{0.03}{0.03} &   --0.21\err{0.05}{0.05}   &      10.5\err{2.1}{2.2}      & \gap{}0.00\err{0.15}{0.18} & --0.31\err{0.06}{0.06} & 1.54 \\
			R = $\frac{2}{3}$\,R$_{\rm eff}$ & 0.10\err{0.05}{0.05} &   --0.21\err{0.08}{0.08}   & \phantom{0}6.7\err{3.9}{1.6} & \gap{}0.00\err{0.24}{0.35} & --0.41\err{0.10}{0.02} & 0.35\\
			R = R$_{\rm eff}$                & 0.00\err{0.07}{0.08} &    --0.21\err{0.1}{0.1}    & \phantom{0}8.0\err{5.6}{2.5} &   --0.36\err{0.43}{0.5}    & --0.43\err{0.14}{0.01} & 0.91 \\ \hline
			best-fit gradients               &  --0.20 $\pm$ 0.01   &     --0.17 $\pm$ 0.02      &       --0.6 $\pm$ 0.8        &     --0.48 $\pm$ 0.07      &    0.00 $\pm$ 0.04     & \\ \hline
		\end{tabular}
	\end{table*}

\begin{figure*}
		\centering
		\includegraphics[width=0.98\textwidth]{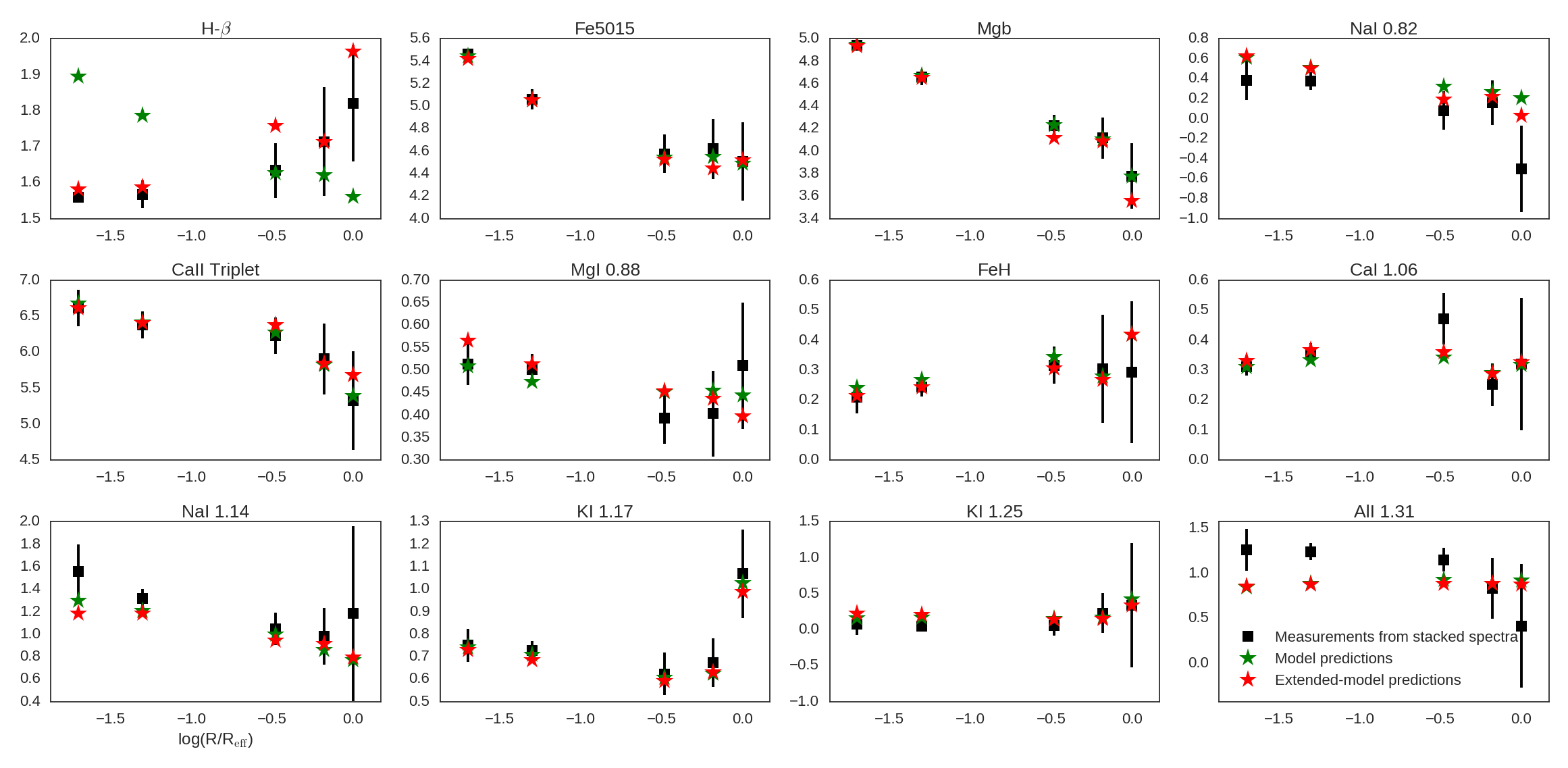}
		\caption{ The results from the fitting procedure applied to the stacked spectra, projected onto the data (black points: equivalent width measurements from the stacked spectra, corrected to 230\,kms$^{-1}$. A green star has been used for the model predictions, while the red stars are the predictions from a different version of the model with additional parameters log(Age) and [C/H]. These allow H$\beta$ to be adequately fit but do not substantially improve the fit to the other lines. The Al\,I 1.31$\umu$m feature was not included in the fit but can be compared with the predictions of the models.}
		\label{fig:bestfit}
	\end{figure*}

\begin{landscape}
		\begin{figure}
			\centering
			\includegraphics[scale=0.42]{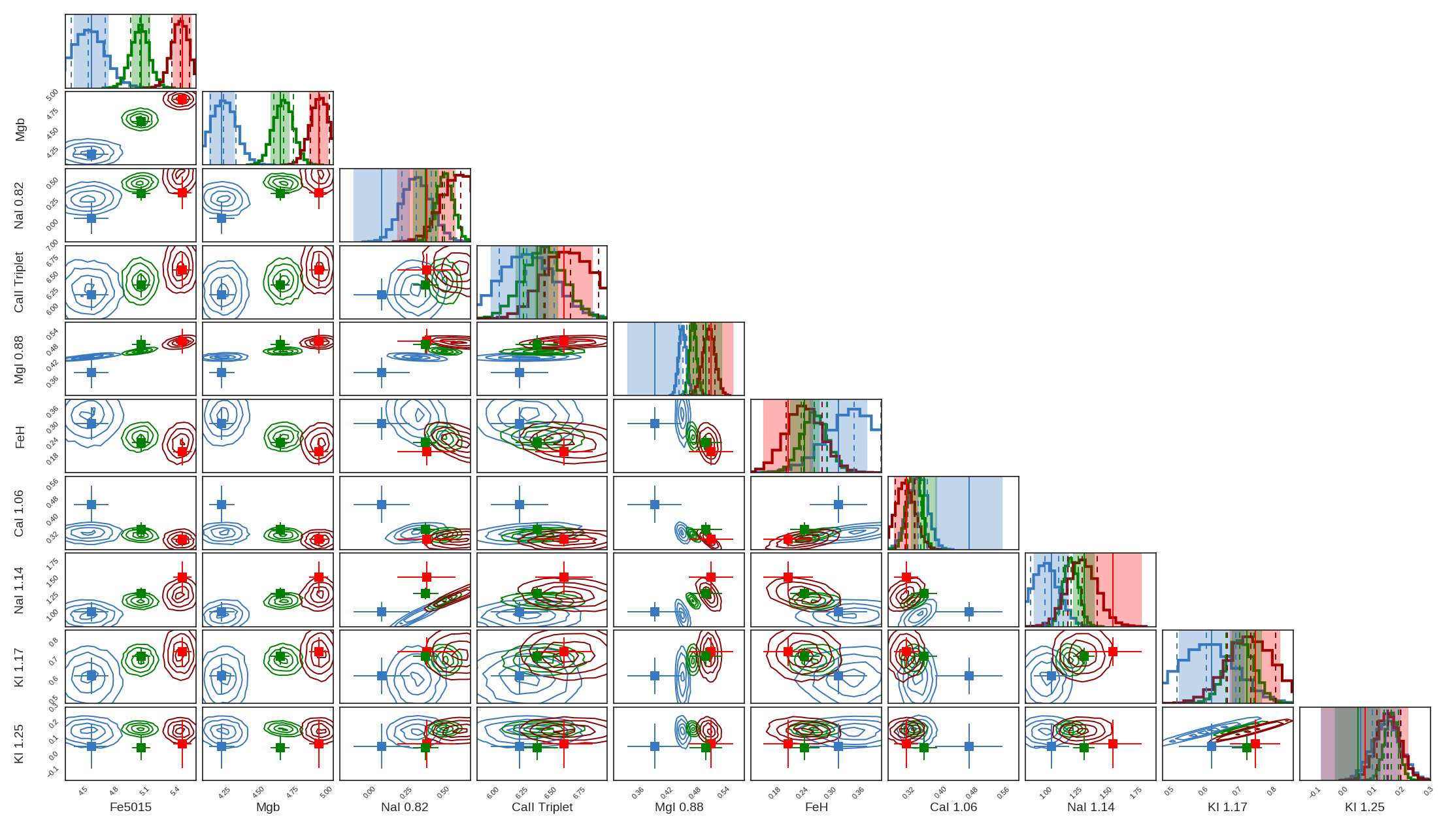}
			\caption{An alternative comparison of the best-fit model predictions and the measurements from the stacked data. The colour-scheme is as in Figure \ref{fig:MCMC_grad}; contours/histograms indicate the distribution of line strengths corresponding to the MCMC parameter sampling distribution while points indicate the data for the three highest S/N stacks.}
			\label{fig:model_draws}
		\end{figure}
	\end{landscape}

	Figure \ref{fig:bestfit} shows the level of agreement between the best fit model and the feature strengths derived from the stacked spectra. We also show the predictions of the model for the data which were not used to directly constrain the fit (H$\beta$ and the Al\,I 1.31$\umu$m doublet). Figure \ref{fig:model_draws} is related, showing the outcome of the modelling process projected onto index-index space (i.e. the distributions of index strengths corresponding to the distributions of parameter values) alongside the data for the three innermost stacked spectra.  We now discuss a few salient features of these plots.

	\subsubsection*{Stellar population age}
	
	In our model prescription we have mostly ignored the effects of stellar population age. NIR features have little sensitivity to this: in the SSP models the strengths of most of the absorption features we measure change by a negligible amount for populations with ages between 7\,Gyr and 13.5\,Gyr. Nevertheless, the inferred IMF does have a dependence on stellar population age, as detailed in Appendix C (dwarf stars contribute less to the total light in a younger population, since slightly higher mass stars will remain on the main sequence). The optical line strengths we use to constrain [$\alpha$/H] and [Fe/H] are also sensitive to age.
	
	However, the optical line strength maps reveal that the stellar populations in our sample are -- as expected for massive ETGs -- uniformly extremely old ($\sim 13.5$\,Gyr) in their cores. In most of these, the H$\beta$ line is weaker than can be readily accounted for within the CvD model grid using the parameters discussed so far. This discrepancy can be reconciled (at the 1$\sigma$ level) by a fairly large enhancement to [C/H], an additional parameter to which our other indices are insensitive. This means that while H$\beta$ can be adequately recovered by the models, doing so requires us to fit for two additional parameters; the consequence is that, without additional constraints on [C/H], stellar age is not well constrained. The results of this test are given in Table \ref{table:carb_age_results}, which shows that marginalising over these two additional parameters does not strongly affect our results. The predictions of this model with extra freedom are shown in Fig. \ref{fig:bestfit} alongside those of the standard model.
	
	\begin{table*}
		\centering
		\caption{ Model parameters and best fit values for stacked spectra at five radii, after marginalising over log(Age) and [C/H]. }
		\label{table:carb_age_results}
		\begin{tabular}{lcccccc}
			\hline
			Extraction Region                &   $\rm [\alpha/H]$   &       $\rm [Fe/H] $        &     f$_{\mathrm{dwarf}}$     &       $\rm [Na/H] $        &   $\rm [Ca/\alpha] $   & reduced $\chi^2$\\ \hline	
			R $< 0.02$\,R$_{\rm eff}$        & 0.39\err{0.03}{0.03} & \gap{}0.14\err{0.03}{0.05} & \phantom{0}8.1\err{2.4}{1.9} &    0.64\err{0.13}{0.12}    & --0.40\err{0.06}{0.03} & 2.30 \\
			R = 0.02--0.05\,R$_{\rm eff}$    & 0.30\err{0.02}{0.03} & \gap{}0.03\err{0.03}{0.04} &      11.8\err{2.5}{1.6}      &    0.51\err{0.05}{0.08}    & --0.34\err{0.04}{0.04} & 3.31 \\
			R = $\frac{1}{3}$\,R$_{\rm eff}$ & 0.16\err{0.03}{0.04} &   --0.13\err{0.05}{0.08}   &      11.0\err{4.0}{2.5}      &    0.07\err{0.14}{0.20}    & --0.31\err{0.06}{0.06} & 2.63 \\
			R = $\frac{2}{3}$\,R$_{\rm eff}$ & 0.11\err{0.06}{0.05} &   --0.23\err{0.11}{0.08}   & \phantom{0}7.0\err{4.9}{1.8} &    0.04\err{0.23}{0.34}    & --0.41\err{0.10}{0.03} & 0.29 \\
			R = R$_{\rm eff}$                & 0.03\err{0.08}{0.07} &   --0.18\err{0.1}{0.13}    & \phantom{0}7.8\err{7.8}{2.3} &   --0.26\err{0.42}{0.51}   & --0.39\err{0.12}{0.04} & 1.57 \\ \hline
			best-fit gradients               &  --0.18 $\pm$ 0.01   &     --0.20 $\pm$ 0.03      &        0.3 $\pm$ 1.5         &     --0.43 $\pm$ 0.06      &    0.01 $\pm$ 0.02     & \\ \hline
		\end{tabular}
	\end{table*}
	
	\subsubsection*{Sodium lines}
	
	A consideration of the fits to the Na\,I features makes clear that the best-fit value of [Na/H] is a compromise between the two, with the solution predicting stronger Na\,I 0.82$\umu$m than is observed and weaker Na\,I 1.14$\umu$m. However, the line strength \textit{gradient} appears to be consistent with the gradient in the model predictions in both cases, indicating that there may be a systematic offset in the strength of one or both lines in the models (consistent with the findings of \citealp{2015MNRAS.454L..71S}) or in the data. This is shown more clearly in the panel of Fig. \ref{fig:model_draws} depicting the joint distribution of model predictions for the two indices, where there is a clear offset from the data.
	
	We find that removal of the Na\,I 0.82$\umu$m line weakens the inferred [Na/H] gradient slightly (--0.46$\pm$0.07), but by less than the 1$\sigma$ uncertainty. However, the inferred absolute value of [Na/H] is stronger by 0.14 for the central spectrum. Removing the Na\,I 1.14$\umu$m line also weakens the inferred [Na/H] gradient slightly (--0.49$\pm$0.07); again, by less than the 1$\sigma$ uncertainty. The inferred absolute value of [Na/H] weakens by 0.09 for the central spectrum.
	
	\subsubsection*{Wing-Ford band}
	
	The rather flat radial behaviour of the Wing-Ford band's strength, alongside its weak absolute strength relative to the fiducial CvD model throughout our sample, adds considerable weight to our results. Nevertheless, removing it from the fit does not qualitatively change the outcome of the analysis: we still recover a radially flat f$_{\rm dwarf}$ with f$_{\rm dwarf}$ = 9.9\err{2.5}{2.1}\% in the innermost stack. The [Na/H] gradient steepens slightly (--0.57$\pm$0.11), but by less than the 1$\sigma$ uncertainty, while the inferred absolute value of [Na/H] decreases by 0.12 for the central spectrum.
	
	\subsubsection*{Potassium lines}
	
	In practice, because an enhancement of [$\alpha$/H] ensures that the K\,I 1.25$\umu$m line becomes extremely weak in the CvD models, we find that including it in our analysis has no substantive effect. This is unfortunate, since the combination of two K\,I features ought to have allowed us to constrain [K/H] while providing additional constraints on the IMF. In reality, [K/H] is effectively constrained only by K\,I 1.17$\umu$m in the high S/N spectra -- and even then, only loosely; e.g. in the stacked spectra from the two innermost regions we fit [K/H] = $-0.12$\err{0.25}{0.25} and $-0.07$\err{0.16}{0.16} Given the uncertainties and because no additional information about the IMF is carried, we cannot conclude much of interest about the behaviour of [K/H] and do not present it in our headline results.
	
	We find that removing the K features and the [K/H] parameter from the fit has a negligible effect on the other parameters.
	
	\subsubsection*{Al\,I 1.31$\umu$m}
	
	The CvD models do not currently incorporate an [Al/H] parameter, so as discussed in section 3.2 we chose not to try and fit this line. However, while the data at the effective radius are consistent with the predictions of our best-fit model, those in the core are in significant tension. At face value this indicates enhanced [Al/H] in the core regions of our sample.
	
	\subsection{Other model parameters}
	
	Including other parameters in the model than those already discussed does not produce a significantly better fit to the data. However, it is important to consider the possible effects of certain `nuisance parameters', which, when marginalised over, may alter the preferred values for the parameters of interest. For this reason we re-analysed our data with two additional parameters included in the model.
	
	The revised CvD models include an `effective temperature' (T$_{\rm eff}$) parameter which encapsulates a shift in the (solar metallicity) isochrones they use. Such shifts can occur given changes in total metallicity, or simply due to uncertainty in the fundamental calibration of the stellar models. For example, \cite{2015ApJ...803...87S} found marginal evidence that in ETGs comparable to our sample the isochrones were shifted to lower effective temperature (--50$\pm$30\,K).
	
	It is important to consider whether $\alpha$-element abundances which are not directly constrained from the data track Mg abundance (as expected). The CvD models allow for simultaneous variation of O, Ne, and S abundances to test this scenario, so we can include in our model a variable [O,Ne,S/$\alpha$] parameter.
	
	These parameters are, as expected, not well constrained by the data. In the stacked spectra we find little evidence for a gradient in T$_{\rm eff}$ (preferred gradient 19$\pm$64K) and find T$_{\rm eff}$ = +58\err{62}{51}K in the central stack. For [O,Ne,S/$\alpha$] we find 0.32\err{0.23}{0.32} in the central stack and a gradient of --0.33$\pm$0.05. This result is driven by the weakness of the K\,I 1.25$\umu$m line, the only feature with measurable sensitivity to the [O,Ne,S/$\alpha$] parameter. The extremely high value inferred is likely non-physical, but does not affect the other spectral features.
	
	Marginalising over these two additional parameters alters the preferred values of the parameters of interest at a fairly minor level, the details of which are contained in Table \ref{table:stackparsandresults2}. In summary: the inferred values for [$\alpha$/H] and [Fe/H] are not significantly changed; the inferred gradients for [Na/H] and [Ca/$\alpha$] are also not significantly changed, however the absolute values for these parameters are reduced by $\sim0.1$ and $\sim$0.05 respectively; we infer a slightly negative f$_{\rm dwarf}$ gradient of --2.6$\pm$1.1\%. These changes are driven by the T$_{\rm eff}$ parameter, with [O,Ne,S/$\alpha$] variations proving inconsequential. A positive T$_{\rm eff}$ weakens the Wing-Ford band slightly, allowing the model freedom to return a combination of lower [Na/H] enhancements and higher f$_{\rm dwarf}$. 
	
	\begin{table*}
		\centering
		\caption{ Model parameters and best fit values for stacked spectra at five radii, after marginalising over T$_{\rm eff}$ and [O,Ne,S/$\alpha$]. }
		\label{table:stackparsandresults2}
		\begin{tabular}{lcccccc}
			\hline
			Extraction Region                &   $\rm [\alpha/H]$   &       $\rm [Fe/H] $        &     f$_{\mathrm{dwarf}}$     &       $\rm [Na/H] $        &   $\rm [Ca/\alpha] $   & reduced $\chi^2$\\ \hline
			R $< 0.02$\,R$_{\rm eff}$        & 0.41\err{0.03}{0.03} & \gap{}0.04\err{0.03}{0.03} &      11.9\err{4.3}{3.6}      & \gap{}0.65\err{0.12}{0.18} & --0.40\err{0.07}{0.03} & 1.49 \\
			R = 0.02--0.05\,R$_{\rm eff}$    & 0.30\err{0.02}{0.02} &   --0.06\err{0.03}{0.03}   &      14.7\err{3.6}{3.7}      & \gap{}0.37\err{0.10}{0.11} & --0.37\err{0.05}{0.04} & 3.24 \\
			R = $\frac{1}{3}$\,R$_{\rm eff}$ & 0.14\err{0.04}{0.04} &   --0.22\err{0.05}{0.05}   &      12.1\err{5.3}{4.5}      &   --0.09\err{0.24}{0.26}   & --0.36\err{0.08}{0.06} & 3.07 \\
			R = $\frac{2}{3}$\,R$_{\rm eff}$ & 0.09\err{0.05}{0.06} &   --0.22\err{0.08}{0.08}   & \phantom{0}6.8\err{5.5}{1.7} & \gap{}0.03\err{0.24}{0.39} & --0.43\err{0.12}{0.01} & 0.37 \\
			R = R$_{\rm eff}$                & 0.01\err{0.08}{0.08} &   --0.24\err{0.11}{0.1}    & \phantom{0}7.7\err{7.7}{2.3} &   --0.39\err{0.45}{0.5}    & --0.43\err{0.14}{0.01} & 1.63 \\ \hline
			best-fit gradients               &  --0.20 $\pm$ 0.01   &     --0.17 $\pm$ 0.02      &       --2.9 $\pm$ 1.1        &     --0.46 $\pm$ 0.08      &   0.00 $\pm$ 0.03    & \\ \hline
		\end{tabular}
	\end{table*}
	
	\section{Discussion}
	
	We have measured average line strength gradients for a stacked sample of eight ETGs. We have investigated nine infrared spectral features, most notably finding a strong radial gradient in the strength of the Na$\,$I 1.14$\umu$m doublet and no significant gradient in the strength of the Wing-Ford molecular band. At an empirical level this is in agreement with the line strength gradients in the samples measured by \cite{2016ApJ...821...39M} and \cite{2016MNRAS.457.1468L}. McConnell et al. argue, through qualitative reference to the CvD models, that the observed gradients are best explained by a large [Na/H] enhancement in the core, rather than by a bottom-heavy IMF concentrated in the core. Our quantitative analysis of the line strengths in the context of the CvD models is in agreement with this, finding evidence for at most only modest radial IMF variation (Chabrier at the effective radius, Salpeter in the innermost region) on average. However, we find that the IMFs of these galaxies are in general bottom-heavy and find evidence for radial IMF variations in some \textit{individual} galaxies.
	
	By contrast, \cite{2016MNRAS.457.1468L} find that their measurements are best accounted for by an IMF gradient, but only if the IMF strongly deviates from a single power law functional form. This is motivated by the measured absence of a radial gradient in the strength of the Wing-Ford band (which we also find) in concert with their measurements of gradients for two TiO indices. A broken power law can reproduce this since these TiO features and the Wing-Ford band are strongest in stars with different ranges of stellar masses ($\lesssim$0.3\,M$_{\odot}$ for the Wing-Ford band, $\sim$0.6\,M$_{\odot}$ for the TiO features). We note that, unlike TiO, the Na$\,$I 1.14$\umu$m feature and the Wing-Ford band are both selectively sensitive to stars of similar masses \citep{2009ApJS..185..289R}, so large deviations from a single power law for the IMF functional form are not a good explanation for our results. Rather, the relative weakness of the Wing-Ford band in the central regions of our sample, despite the enhanced [Fe/H], can best be explained by the high sodium enrichment there (a possibility not accounted for in the method of La Barbera et al.), since Na is an electron donor and promotes the dissociation of the FeH molecule.
	
	The index gradients we measure are also qualitatively similar to those measured in M31 by \cite{2015MNRAS.452..597Z}. Whilst the authors do not explicitly calculate gradients, they find the Wing-Ford band does not vary in strength significantly with radius in M31. Meanwhile, a steep Na\,I 0.82$\umu$m gradient is measured within the central $\sim40\arcsec$ (corresponding to $<0.1$R$_{\rm eff}$ in our sample). The authors show that a consideration of these two features with respect to the CvD12a model grid favours no deviation from a Milky-Way-like IMF throughout the galaxy, with an enhancement of +0.7 in [Na/H] in the core relative to the disc. In CvD12b a spectral fit indicated an only moderately bottom-heavy IMF in the bulge of M31, which was in fact consistent with the results of Zieleniewski et al. The gradients we measure in ETGs are hence at least qualitatively similar to those measured in a galaxy not thought to host a stellar population formed according to a notably bottom-heavy IMF.
	
	Since the submission of this paper a number of other works were published on the subject of radial IMF gradients. \cite{2016arXiv161109859V} find evidence for strong radial IMF variations in some members of a sample of six galaxies. By contrast, \cite{2016arXiv161200364V} (three galaxies) and \cite{2017MNRAS.465..192Z} (two galaxies) do not find clear evidence for variations.
	
	Regardless of the variation of the IMF, our results indicate a radial variation in sodium abundance ([Na/H]). The absolute strength of the Na$\,$I 1.14$\umu$m feature is very high in the central extraction region and in tension with the absolute strength of the 0.82$\umu$m feature. This is similar to the result reported in \cite{2015MNRAS.454L..71S} in which this feature was found to be stronger than expected given other constraints on stellar populations. Additionally, as shown in the same work there is a weak sodium line at 1.27$\umu$m, which becomes visible only in model spectra with very high [Na/H] and which is rather insensitive to the IMF (in contrast to the 0.82$\umu$m and 1.14$\umu$m lines). This feature is clearly visible in our stacked galaxy core spectra. For this line we define an absorption index with blue continuum 12655--12665\AA, red continuum 12705-12715\AA, and feature band 12675-12690\AA. Under this definition the fiducial SSP model predicts an equivalent width of 0.14\AA, while an extremely sodium enhanced model ([Na/H]=+0.9) instead predicts 0.27\AA. By contrast, even for an X=3 power law IMF the line's strength is only 0.17\AA. Notably, we measure an equivalent width of 0.33$\pm$ 0.08\AA$\phantom{ }$ in the innermost stacked spectrum. The line is however in a region of particularly strong atmospheric absorption, which becomes prohibitive in the outer rings. Though we do not, therefore, include this feature in our quantitative analysis, it appears to support extreme sodium abundances of [Na/H]$>$+0.9 in the core and is consistent with the [Na/H] we infer from other lines.
	
	The Al$\,$I 1.31$\umu$m line has not been studied in ETGs before. Like Na\,I 1.14$\umu$m it is dramatically stronger than in the models in the cores of our sample of galaxies and cannot be matched even by SSP models with very bottom-heavy IMFs. We note that the galaxies with strong Al$\,$I 1.31$\umu$m features in their core spectra are generally the ones with strong Na$\,$I 1.14$\umu$m features too. Interestingly, sodium and aluminium have closely related formation channels (see \citealp{2007A&A...465..799L}, and references therein). These elements are unusual in having strongly metallicity-dependent Type II SNe yields \citep{2006ApJ...653.1145K}. The Al$\,$I 1.31$\umu$m and Na\,I 1.14$\umu$m features are both sensitive to IMF (in the same sense), but probe slightly different stellar mass ranges (\citealp{2009ApJS..185..289R}; \citealp{2012ApJ...747...69C}): a steeper low-mass slope has a larger effect on the sodium feature's strength than it does on the aluminium feature, which is chiefly sensitive to 0.4--0.7M$_{\odot}$ stars. The extreme strength of the Al$\,$I 1.31$\umu$m feature therefore may be interpreted as further anecdotal support for the hypothesis that chemical abundance effects are the dominant cause of the measured radial variation in absorption feature strengths in ETGs.
	
	Radial gradients in the relative contribution of dwarf stars to the integrated light of ETGs are motivated by a combination of observations and theory, as discussed in the Introduction. Spectroscopy of the cores of ETGs has indicated that the IMF is bottom-heavy in the most massive systems and Milky-Way-like in less massive galaxies; this is supported by measurements of the M/L ratio for these systems via dynamics and lensing. In our work we replicate these results using a set of infrared spectral indices, some of which are not heretofore well-studied in ETGs. Yet our current understanding of the formation of the most massive ETGs is that their outer regions are assembled by accretion of lower mass galaxies through minor mergers, which is required to explain their evolution in size for z$<$2 (e.g. \citealp{2009ApJ...697.1290B}, \citealp{2009MNRAS.398..898H}, \citealp{2010MNRAS.401.1099H}) and supported by simulations (e.g. \citealp{2010ApJ...725.2312O}, \citealp{2012ApJ...744...63O}, \citealp{2013MNRAS.429.2924H}). These merging systems presumably contain stars which formed according to a Milky-Way-like IMF, and they are expected to be deposited at around the effective radius, so this type of growth should lead to a radial gradient in the population dwarf fraction (note, however, that it is believed that ETGs that are slow rotators -- there are several in our sample -- are more likely to have undergone recent major mergers; see \citealp{2010MNRAS.406.2405B}, \citealp{2011MNRAS.417..845K}, and \citealp{2014MNRAS.444.3357N}. These events are much more disruptive. We do not, however, see a clear dichotomy between fast and slow rotators in our results). 
	
	Our work suggests that radial IMF variations in ETGs are on the whole rather gentle, even in those galaxies for which we infer very bottom-heavy central IMFs. In the case of those individual galaxies in our sample for which we are able to make robust estimates of the IMF, most are at least marginally consistent with a non-radially varying IMF. The average behaviour (as probed via the stacked spectra) of our entire sample tracks this, being consistent with a modestly bottom-heavy IMF with a slope slightly steeper than Salpeter at all radii. Within the `inside-out' paradigm of galaxy formation, this would suggest that minor mergers deposit stellar mass largely beyond the effective radius. This possibility has some observational support, with \cite{2010MNRAS.407L..26C} arguing that spectroscopy of the very massive galaxy NGC\,4889 ($\sigma \sim$ 300\,kms$^{-1}$) indicates the deposition of accreted mass at $\gtrsim$1.2\,R$_{\rm eff}$. Meanwhile simulations present an ambiguous picture, with \cite{2016MNRAS.458.2371R} finding that the fractional radius at which accreted stars are deposited varies quite strongly with total stellar mass (being typically less than R$_{\rm eff}$ for the most massive ellipticals, however). The strong average line strength gradients we observe for our sample indicate that the stellar populations at R$_{\rm eff}$ are chemically distinct from those in the core, but do not necessarily show that they formed entirely \textit{ex situ}: such gradients are also a consequence of monolithic collapse (see e.g. \citealp{2004MNRAS.347..740K}). A shallow IMF gradient could indicate the following: first, that the IMF is uniform within the progenitor systems that form the cores of present day ETGs (being bottom-heavy throughout) and secondly, that minor systems accreted onto this core deposit little mass inside the effective radius. Thus, our results may clarify the details of this growth channel by constraining the radius at which most of the accreted mass is deposited.
	
	\section{Conclusions}
	
	In this work we have studied the spatially-resolved infrared spectra of eight nearby ETGs, extracting information about the radial variation in the strength of a variety of absorption features between 0.8 and 1.35$\umu$m, including some which have not been previously studied. These were chosen for their sensitivity to both the IMF and a variety of chemical abundances. By interpreting the strength of these features with reference to the CvD stellar population models, we probe the radial variation of the stellar populations of our sample of galaxies. Our main conclusions are as follows:

	\begin{enumerate}
		\item In stacked spectra derived from our sample we measure significant negative gradients in the strengths of several IMF-sensitive features, namely the Na\,I 0.82 and 1.14$\umu$m features (--0.34$\pm$0.05\AA\, and --0.40$\pm$0.05\AA\, per decade in R/R$_{\rm eff}$, respectively), the Ca\,II Triplet (--0.45$\pm$0.14\AA), the Mg\,I 0.88$\umu$m line (-0.08$\pm$0.02\AA) and Al\,I 1.31$\umu$m doublet (--0.34$\pm$0.12\AA). We measure a marginal positive gradient for the Wing-Ford band (+0.06$\pm$0.04\AA), while the Ca\,I 1.03$\umu$m line, K\,I 1.17$\umu$m composite feature, and the K\,I 1.25$\umu$m line are also consistent with no radial variation. The only IMF-sensitive features in which strong radial absorption strength gradients are found are those for which enhancing the stellar population dwarf-fraction is degenerate with enhancing the abundance of sodium or aluminium (which are formed by related nucleosynthetic mechanisms).	
		\item We interpret these results in terms of variations in both the dwarf star content and element abundances using state-of-the-art stellar population models and Markov-Chain Monte Carlo parameter estimation. The absorption index gradients we measure can be primarily explained by radial variations in the chemical abundance pattern of the stellar population. We infer from the stacked spectra an average sodium abundance (i.e. [Na/H]) gradient of --0.53 $\pm$ 0.07 per decade in radius (alternatively expressed as a [Na/Fe] gradient of $-0.35$), leading to an enhancement of [Na/H]\,$\gtrsim$\,$+0.7$ in the core. This may be connected to the metallicity-dependent yield of sodium in Type II SNe. We infer an average gradient in the fractional contribution of dwarf stars to the  luminosity, $f_{\rm dwarf}$, of --0.7$\pm$0.7\% (the difference in f$_{\rm dwarf}$ between Chabrier and Salpeter is $\sim 4\%$).	
		\item The chemical abundance gradients for individual galaxies in our ETG sample (those with the highest quality, most comprehensive data) appear comparable to those derived from the stacked spectra. These gradients appear to be fairly uniform across the sample, even though the chemical abundances themselves vary substantially within the sample at fixed radius.
		\item Our results suggest that on average the IMF in our sample galaxies is bottom-heavy out to a significant fraction of the effective radius, with little evidence for a strong gradient. There may however be significant variation in radial IMF trends between galaxies. Our results may suggest that minor mergers deposit stars largely beyond the effective radius of the present-day systems.
	\end{enumerate}
	
	Our results will be supplemented in the near future by newly-acquired KMOS observations of our sample in the K-band. This band contains a variety of strong, gravity-sensitive absorption features associated with a variety of elements. These features are largely unexplored in recent work (but see \citealp{2008ApJ...674..194S}) and should help us further constrain the stellar population parameters of ETGs. In particular, the Na$\,$I 2.21$\umu$m feature is readily measurable and, like the weak 1.27$\umu$m line, is relatively insensitive to the IMF. Conversely, other K-band features (such as the Ca\,I line at 1.98$\umu$m and the CO-bandheads at 2.30$\umu$m) are very insensitive to [Na/H] but are gravity-sensitive. These observations should therefore allow much tighter constraints to be placed on both sodium abundance and any IMF variation. 
	
	In future work we will investigate whether radial gradients in stellar population parameters are linked to the \textit{global} properties of ETGs (such as mass, velocity dispersion etc.) and hence their formation histories. This would be challenging with the data presented here alone. We intend to study global trends (variation across a sample of galaxies at fixed radius) with a larger sample of cluster ETGs at z$\approx$0.05 using KMOS data covering the same wavelength ranges as in this work. A more sophisticated analysis taking full account of both radial gradients and population trends simultaneously will enable us to better understand the variation of a variety of stellar population properties, including the IMF, within and between galaxies.

	\section*{Acknowledgements}
	
	The authors are grateful to Ray Sharples for the use of KMOS GTO time for some observations and for useful discussions regarding instrumentation; our thanks also to Renske Smit and Helen Johnson for obtaining additional observations of NGC\,1407 and NGC\,3379 for this project. We wish to thank Charlie Conroy for providing up-to-date versions of the CvD models, and the anonymous referee for helpful comments which improved the paper. 
	
	PA was supported by an STFC studentship (ST/K501979/1) and RJS by the STFC Durham Astronomy Consolidated Grant 2014--2017 (ST/L00075X/1). The data used here are available through the ESO science archive (programme IDs: 093.B-0693(A), 094.B-0061(C)). 
	
	\bibliographystyle{mn2e}
	\bibliography{PAlton}

\begin{thebibliography}{}
\expandafter\ifx\csname natexlab\endcsname\relax\def\natexlab#1{#1}\fi

\bibitem[{{Auger} {et~al.}(2010){Auger}, {Treu}, {Gavazzi}, {Bolton},
  {Koopmans}, \& {Marshall}}]{2010ApJ...721L.163A}
{Auger}, M.~W., {Treu}, T., {Gavazzi}, R., {et~al.} 2010, \apjl, 721, L163

\bibitem[{{Bastian} {et~al.}(2010){Bastian}, {Covey}, \&
  {Meyer}}]{2010ARA&A..48..339B}
{Bastian}, N., {Covey}, K.~R., \& {Meyer}, M.~R. 2010, \araa, 48, 339

\bibitem[{{Bezanson} {et~al.}(2009){Bezanson}, {van Dokkum}, {Tal},
  {Marchesini}, {Kriek}, {Franx}, \& {Coppi}}]{2009ApJ...697.1290B}
{Bezanson}, R., {van Dokkum}, P.~G., {Tal}, T., {et~al.} 2009, \apj, 697, 1290

\bibitem[{{Bois} {et~al.}(2010){Bois}, {Bournaud}, {Emsellem}, {Alatalo},
  {Blitz}, {Bureau}, {Cappellari}, {Davies}, {Davis}, {de Zeeuw}, {Duc},
  {Khochfar}, {Krajnovi{\'c}}, {Kuntschner}, {Lablanche}, {McDermid},
  {Morganti}, {Naab}, {Oosterloo}, {Sarzi}, {Scott}, {Serra}, {Weijmans}, \&
  {Young}}]{2010MNRAS.406.2405B}
{Bois}, M., {Bournaud}, F., {Emsellem}, E., {et~al.} 2010, \mnras, 406, 2405

\bibitem[{{Cappellari} \& {Emsellem}(2004)}]{2004PASP..116..138C}
{Cappellari}, M., \& {Emsellem}, E. 2004, \pasp, 116, 138

\bibitem[{{Cappellari} {et~al.}(2011){Cappellari}, {Emsellem}, {Krajnovi{\'c}},
  {McDermid}, {Scott}, {Verdoes Kleijn}, {Young}, {Alatalo}, {Bacon}, {Blitz},
  {Bois}, {Bournaud}, {Bureau}, {Davies}, {Davis}, {de Zeeuw}, {Duc},
  {Khochfar}, {Kuntschner}, {Lablanche}, {Morganti}, {Naab}, {Oosterloo},
  {Sarzi}, {Serra}, \& {Weijmans}}]{2011MNRAS.413..813C}
{Cappellari}, M., {Emsellem}, E., {Krajnovi{\'c}}, D., {et~al.} 2011, \mnras,
  413, 813

\bibitem[{{Cappellari} {et~al.}(2012){Cappellari}, {McDermid}, {Alatalo},
  {Blitz}, {Bois}, {Bournaud}, {Bureau}, {Crocker}, {Davies}, {Davis}, {de
  Zeeuw}, {Duc}, {Emsellem}, {Khochfar}, {Krajnovi{\'c}}, {Kuntschner},
  {Lablanche}, {Morganti}, {Naab}, {Oosterloo}, {Sarzi}, {Scott}, {Serra},
  {Weijmans}, \& {Young}}]{2012Natur.484..485C}
{Cappellari}, M., {McDermid}, R.~M., {Alatalo}, K., {et~al.} 2012, \nat, 484,
  485

\bibitem[{{Carter} {et~al.}(1986){Carter}, {Visvanathan}, \&
  {Pickles}}]{1986ApJ...311..637C}
{Carter}, D., {Visvanathan}, N., \& {Pickles}, A.~J. 1986, \apj, 311, 637

\bibitem[{{Chabrier}(2003)}]{2003PASP..115..763C}
{Chabrier}, G. 2003, \pasp, 115, 763

\bibitem[{{Chabrier} {et~al.}(2014){Chabrier}, {Hennebelle}, \&
  {Charlot}}]{2014ApJ...796...75C}
{Chabrier}, G., {Hennebelle}, P., \& {Charlot}, S. 2014, \apj, 796, 75

\bibitem[{{Coccato} {et~al.}(2010){Coccato}, {Gerhard}, \&
  {Arnaboldi}}]{2010MNRAS.407L..26C}
{Coccato}, L., {Gerhard}, O., \& {Arnaboldi}, M. 2010, \mnras, 407, L26

\bibitem[{{Cohen}(1979)}]{1979ApJ...228..405C}
{Cohen}, J.~G. 1979, \apj, 228, 405

\bibitem[{{Conroy} {et~al.}(2014){Conroy}, {Graves}, \& {van
  Dokkum}}]{2014ApJ...780...33C}
{Conroy}, C., {Graves}, G.~J., \& {van Dokkum}, P.~G. 2014, \apj, 780, 33

\bibitem[{{Conroy} \& {van Dokkum}(2012{\natexlab{a}})}]{2012ApJ...747...69C}
{Conroy}, C., \& {van Dokkum}, P. 2012{\natexlab{a}}, \apj, 747, 69

\bibitem[{{Conroy} \& {van Dokkum}(2012{\natexlab{b}})}]{2012ApJ...760...71C}
{Conroy}, C., \& {van Dokkum}, P.~G. 2012{\natexlab{b}}, \apj, 760, 71

\bibitem[{{Davies}(2007)}]{2007MNRAS.375.1099D}
{Davies}, R.~I. 2007, \mnras, 375, 1099

\bibitem[{{Emsellem} {et~al.}(2004){Emsellem}, {Cappellari}, {Peletier},
  {McDermid}, {Bacon}, {Bureau}, {Copin}, {Davies}, {Krajnovi{\'c}},
  {Kuntschner}, {Miller}, \& {de Zeeuw}}]{2004MNRAS.352..721E}
{Emsellem}, E., {Cappellari}, M., {Peletier}, R.~F., {et~al.} 2004, \mnras,
  352, 721

\bibitem[{{Foreman-Mackey} {et~al.}(2013){Foreman-Mackey}, {Hogg}, {Lang}, \&
  {Goodman}}]{2013PASP..125..306F}
{Foreman-Mackey}, D., {Hogg}, D.~W., {Lang}, D., \& {Goodman}, J. 2013, \pasp,
  125, 306

\bibitem[{{Goodman} \& {Weare}(2010)}]{Goodman_and_Weare}
{Goodman}, J., \& {Weare}, J. 2010, CAMCoS, 5, 65

\bibitem[{{Hilz} {et~al.}(2013){Hilz}, {Naab}, \&
  {Ostriker}}]{2013MNRAS.429.2924H}
{Hilz}, M., {Naab}, T., \& {Ostriker}, J.~P. 2013, \mnras, 429, 2924

\bibitem[{{Ho} {et~al.}(1997){Ho}, {Filippenko}, \&
  {Sargent}}]{1997ApJS..112..315H}
{Ho}, L.~C., {Filippenko}, A.~V., \& {Sargent}, W.~L.~W. 1997, \apjs, 112, 315

\bibitem[{{Hopkins} {et~al.}(2010){Hopkins}, {Bundy}, {Hernquist}, {Wuyts}, \&
  {Cox}}]{2010MNRAS.401.1099H}
{Hopkins}, P.~F., {Bundy}, K., {Hernquist}, L., {Wuyts}, S., \& {Cox}, T.~J.
  2010, \mnras, 401, 1099

\bibitem[{{Hopkins} {et~al.}(2009){Hopkins}, {Bundy}, {Murray}, {Quataert},
  {Lauer}, \& {Ma}}]{2009MNRAS.398..898H}
{Hopkins}, P.~F., {Bundy}, K., {Murray}, N., {et~al.} 2009, \mnras, 398, 898

\bibitem[{{Jarrett} {et~al.}(2003){Jarrett}, {Chester}, {Cutri}, {Schneider},
  \& {Huchra}}]{2003AJ....125..525J}
{Jarrett}, T.~H., {Chester}, T., {Cutri}, R., {Schneider}, S.~E., \& {Huchra},
  J.~P. 2003, \aj, 125, 525

\bibitem[{{Kausch} {et~al.}(2015){Kausch}, {Noll}, {Smette}, {Kimeswenger},
  {Barden}, {Szyszka}, {Jones}, {Sana}, {Horst}, \&
  {Kerber}}]{2015A&A...576A..78K}
{Kausch}, W., {Noll}, S., {Smette}, A., {et~al.} 2015, \aap, 576, A78

\bibitem[{{Khochfar} {et~al.}(2011){Khochfar}, {Emsellem}, {Serra}, {Bois},
  {Alatalo}, {Bacon}, {Blitz}, {Bournaud}, {Bureau}, {Cappellari}, {Davies},
  {Davis}, {de Zeeuw}, {Duc}, {Krajnovi{\'c}}, {Kuntschner}, {Lablanche},
  {McDermid}, {Morganti}, {Naab}, {Oosterloo}, {Sarzi}, {Scott}, {Weijmans}, \&
  {Young}}]{2011MNRAS.417..845K}
{Khochfar}, S., {Emsellem}, E., {Serra}, P., {et~al.} 2011, \mnras, 417, 845

\bibitem[{{Kobayashi}(2004)}]{2004MNRAS.347..740K}
{Kobayashi}, C. 2004, \mnras, 347, 740

\bibitem[{{Kobayashi} {et~al.}(2006){Kobayashi}, {Umeda}, {Nomoto}, {Tominaga},
  \& {Ohkubo}}]{2006ApJ...653.1145K}
{Kobayashi}, C., {Umeda}, H., {Nomoto}, K., {Tominaga}, N., \& {Ohkubo}, T.
  2006, \apj, 653, 1145

\bibitem[{{Kroupa}(2001)}]{2001MNRAS.322..231K}
{Kroupa}, P. 2001, \mnras, 322, 231

\bibitem[{{La Barbera} {et~al.}(2013){La Barbera}, {Ferreras}, {Vazdekis}, {de
  la Rosa}, {de Carvalho}, {Trevisan}, {Falc{\'o}n-Barroso}, \&
  {Ricciardelli}}]{2013MNRAS.433.3017L}
{La Barbera}, F., {Ferreras}, I., {Vazdekis}, A., {et~al.} 2013, \mnras, 433,
  3017

\bibitem[{{La Barbera} {et~al.}(2016){La Barbera}, {Vazdekis}, {Ferreras},
  {Pasquali}, {Cappellari}, {Mart{\'{\i}}n-Navarro}, {Sch{\"o}nebeck}, \&
  {Falc{\'o}n-Barroso}}]{2016MNRAS.457.1468L}
{La Barbera}, F., {Vazdekis}, A., {Ferreras}, I., {et~al.} 2016, \mnras, 457,
  1468

\bibitem[{{Landt} {et~al.}(2008){Landt}, {Bentz}, {Ward}, {Elvis}, {Peterson},
  {Korista}, \& {Karovska}}]{2008ApJS..174..282L}
{Landt}, H., {Bentz}, M.~C., {Ward}, M.~J., {et~al.} 2008, \apjs, 174, 282

\bibitem[{{Lecureur} {et~al.}(2007){Lecureur}, {Hill}, {Zoccali}, {Barbuy},
  {G{\'o}mez}, {Minniti}, {Ortolani}, \& {Renzini}}]{2007A&A...465..799L}
{Lecureur}, A., {Hill}, V., {Zoccali}, M., {et~al.} 2007, \aap, 465, 799

\bibitem[{{Luhman} {et~al.}(2003){Luhman}, {Brice{\~n}o}, {Stauffer},
  {Hartmann}, {Barrado y Navascu{\'e}s}, \& {Caldwell}}]{2003ApJ...590..348L}
{Luhman}, K.~L., {Brice{\~n}o}, C., {Stauffer}, J.~R., {et~al.} 2003, \apj,
  590, 348

\bibitem[{{Marks} {et~al.}(2012){Marks}, {Kroupa}, {Dabringhausen}, \&
  {Pawlowski}}]{2012MNRAS.422.2246M}
{Marks}, M., {Kroupa}, P., {Dabringhausen}, J., \& {Pawlowski}, M.~S. 2012,
  \mnras, 422, 2246

\bibitem[{{Mart{\'{\i}}n-Navarro} {et~al.}(2015){Mart{\'{\i}}n-Navarro},
  {Barbera}, {Vazdekis}, {Falc{\'o}n-Barroso}, \&
  {Ferreras}}]{2015MNRAS.447.1033M}
{Mart{\'{\i}}n-Navarro}, I., {Barbera}, F.~L., {Vazdekis}, A.,
  {Falc{\'o}n-Barroso}, J., \& {Ferreras}, I. 2015, \mnras, 447, 1033

\bibitem[{{McConnell} {et~al.}(2016){McConnell}, {Lu}, \&
  {Mann}}]{2016ApJ...821...39M}
{McConnell}, N.~J., {Lu}, J.~R., \& {Mann}, A.~W. 2016, \apj, 821, 39

\bibitem[{{McDermid} {et~al.}(2015){McDermid}, {Alatalo}, {Blitz}, {Bournaud},
  {Bureau}, {Cappellari}, {Crocker}, {Davies}, {Davis}, {de Zeeuw}, {Duc},
  {Emsellem}, {Khochfar}, {Krajnovi{\'c}}, {Kuntschner}, {Morganti}, {Naab},
  {Oosterloo}, {Sarzi}, {Scott}, {Serra}, {Weijmans}, \&
  {Young}}]{2015MNRAS.448.3484M}
{McDermid}, R.~M., {Alatalo}, K., {Blitz}, L., {et~al.} 2015, \mnras, 448, 3484

\bibitem[{{Naab} {et~al.}(2014){Naab}, {Oser}, {Emsellem}, {Cappellari},
  {Krajnovi{\'c}}, {McDermid}, {Alatalo}, {Bayet}, {Blitz}, {Bois}, {Bournaud},
  {Bureau}, {Crocker}, {Davies}, {Davis}, {de Zeeuw}, {Duc}, {Hirschmann},
  {Johansson}, {Khochfar}, {Kuntschner}, {Morganti}, {Oosterloo}, {Sarzi},
  {Scott}, {Serra}, {Ven}, {Weijmans}, \& {Young}}]{2014MNRAS.444.3357N}
{Naab}, T., {Oser}, L., {Emsellem}, E., {et~al.} 2014, \mnras, 444, 3357

\bibitem[{{Noll} {et~al.}(2014){Noll}, {Kausch}, {Kimeswenger}, {Barden},
  {Jones}, {Modigliani}, {Szyszka}, \& {Taylor}}]{2014A&A...567A..25N}
{Noll}, S., {Kausch}, W., {Kimeswenger}, S., {et~al.} 2014, \aap, 567, A25

\bibitem[{{Nordh} {et~al.}(1977){Nordh}, {Lindgren}, \&
  {Wing}}]{1977A&A....56....1N}
{Nordh}, H.~L., {Lindgren}, B., \& {Wing}, R.~F. 1977, \aap, 56, 1

\bibitem[{{Oser} {et~al.}(2012){Oser}, {Naab}, {Ostriker}, \&
  {Johansson}}]{2012ApJ...744...63O}
{Oser}, L., {Naab}, T., {Ostriker}, J.~P., \& {Johansson}, P.~H. 2012, \apj,
  744, 63

\bibitem[{{Oser} {et~al.}(2010){Oser}, {Ostriker}, {Naab}, {Johansson}, \&
  {Burkert}}]{2010ApJ...725.2312O}
{Oser}, L., {Ostriker}, J.~P., {Naab}, T., {Johansson}, P.~H., \& {Burkert}, A.
  2010, \apj, 725, 2312

\bibitem[{{Pietrinferni} {et~al.}(2013){Pietrinferni}, {Cassisi}, {Salaris}, \&
  {Hidalgo}}]{2013A&A...558A..46P}
{Pietrinferni}, A., {Cassisi}, S., {Salaris}, M., \& {Hidalgo}, S. 2013, \aap,
  558, A46

\bibitem[{{Prieto} {et~al.}(2016){Prieto}, {Fern{\'a}ndez-Ontiveros},
  {Markoff}, {Espada}, \& {Gonz{\'a}lez-Mart{\'{\i}}n}}]{2016MNRAS.457.3801P}
{Prieto}, M.~A., {Fern{\'a}ndez-Ontiveros}, J.~A., {Markoff}, S., {Espada}, D.,
  \& {Gonz{\'a}lez-Mart{\'{\i}}n}, O. 2016, \mnras, 457, 3801

\bibitem[{{Rayner} {et~al.}(2009){Rayner}, {Cushing}, \&
  {Vacca}}]{2009ApJS..185..289R}
{Rayner}, J.~T., {Cushing}, M.~C., \& {Vacca}, W.~D. 2009, \apjs, 185, 289

\bibitem[{{Rodriguez-Gomez} {et~al.}(2016){Rodriguez-Gomez}, {Pillepich},
  {Sales}, {Genel}, {Vogelsberger}, {Zhu}, {Wellons}, {Nelson}, {Torrey},
  {Springel}, {Ma}, \& {Hernquist}}]{2016MNRAS.458.2371R}
{Rodriguez-Gomez}, V., {Pillepich}, A., {Sales}, L.~V., {et~al.} 2016, \mnras,
  458, 2371

\bibitem[{{Salpeter}(1955)}]{1955ApJ...121..161S}
{Salpeter}, E.~E. 1955, \apj, 121, 161

\bibitem[{{Sharples} {et~al.}(2013){Sharples}, {Bender}, {Agudo Berbel},
  {Bezawada}, {Castillo}, {Cirasuolo}, {Davidson}, {Davies}, {Dubbeldam},
  {Fairley}, {Finger}, {F{\"o}rster Schreiber}, {Gonte}, {Hess}, {Jung},
  {Lewis}, {Lizon}, {Muschielok}, {Pasquini}, {Pirard}, {Popovic}, {Ramsay},
  {Rees}, {Richter}, {Riquelme}, {Rodrigues}, {Saviane}, {Schlichter},
  {Schmidtobreick}, {Segovia}, {Smette}, {Szeifert}, {van Kesteren}, {Wegner},
  \& {Wiezorrek}}]{2013Msngr.151...21S}
{Sharples}, R., {Bender}, R., {Agudo Berbel}, A., {et~al.} 2013, The Messenger,
  151, 21

\bibitem[{{Silva} {et~al.}(2008){Silva}, {Kuntschner}, \&
  {Lyubenova}}]{2008ApJ...674..194S}
{Silva}, D.~R., {Kuntschner}, H., \& {Lyubenova}, M. 2008, \apj, 674, 194

\bibitem[{{Smette} {et~al.}(2015){Smette}, {Sana}, {Noll}, {Horst}, {Kausch},
  {Kimeswenger}, {Barden}, {Szyszka}, {Jones}, {Gallenne}, {Vinther},
  {Ballester}, \& {Taylor}}]{2015A&A...576A..77S}
{Smette}, A., {Sana}, H., {Noll}, S., {et~al.} 2015, \aap, 576, A77

\bibitem[{{Smith}(2014)}]{2014MNRAS.443L..69S}
{Smith}, R.~J. 2014, \mnras, 443, L69

\bibitem[{{Smith} {et~al.}(2015){Smith}, {Alton}, {Lucey}, {Conroy}, \&
  {Carter}}]{2015MNRAS.454L..71S}
{Smith}, R.~J., {Alton}, P., {Lucey}, J.~R., {Conroy}, C., \& {Carter}, D.
  2015, \mnras, 454, L71

\bibitem[{{Smith} {et~al.}(2012){Smith}, {Lucey}, \&
  {Carter}}]{2012MNRAS.426.2994S}
{Smith}, R.~J., {Lucey}, J.~R., \& {Carter}, D. 2012, \mnras, 426, 2994

\bibitem[{{Spiniello} {et~al.}(2015){Spiniello}, {Trager}, \&
  {Koopmans}}]{2015ApJ...803...87S}
{Spiniello}, C., {Trager}, S.~C., \& {Koopmans}, L.~V.~E. 2015, \apj, 803, 87

\bibitem[{{Spiniello} {et~al.}(2012){Spiniello}, {Trager}, {Koopmans}, \&
  {Chen}}]{2012ApJ...753L..32S}
{Spiniello}, C., {Trager}, S.~C., {Koopmans}, L.~V.~E., \& {Chen}, Y.~P. 2012,
  \apjl, 753, L32

\bibitem[{{Spinrad} \& {Taylor}(1971)}]{1971ApJS...22..445S}
{Spinrad}, H., \& {Taylor}, B.~J. 1971, \apjs, 22, 445

\bibitem[{{Thomas} {et~al.}(2003){Thomas}, {Maraston}, \&
  {Bender}}]{2003MNRAS.343..279T}
{Thomas}, D., {Maraston}, C., \& {Bender}, R. 2003, \mnras, 343, 279

\bibitem[{{Thomas} {et~al.}(2011){Thomas}, {Saglia}, {Bender}, {Thomas},
  {Gebhardt}, {Magorrian}, {Corsini}, {Wegner}, \&
  {Seitz}}]{2011MNRAS.415..545T}
{Thomas}, J., {Saglia}, R.~P., {Bender}, R., {et~al.} 2011, \mnras, 415, 545

\bibitem[{{van Dokkum} {et~al.}(2016){van Dokkum}, {Conroy}, {Villaume},
  {Brodie}, \& {Romanowsky}}]{2016arXiv161109859V}
{van Dokkum}, P., {Conroy}, C., {Villaume}, A., {Brodie}, J., \& {Romanowsky},
  A. 2016, ArXiv e-prints, arXiv:1611.09859

\bibitem[{{van Dokkum} \& {Conroy}(2010)}]{2010Natur.468..940V}
{van Dokkum}, P.~G., \& {Conroy}, C. 2010, \nat, 468, 940

\bibitem[{{Vaughan} {et~al.}(2016){Vaughan}, {Houghton}, {Davies}, \&
  {Zieleniewski}}]{2016arXiv161200364V}
{Vaughan}, S.~P., {Houghton}, R.~C.~W., {Davies}, R.~L., \& {Zieleniewski}, S.
  2016, ArXiv e-prints, arXiv:1612.00364

\bibitem[{{Vazdekis} {et~al.}(1996){Vazdekis}, {Casuso}, {Peletier}, \&
  {Beckman}}]{1996ApJS..106..307V}
{Vazdekis}, A., {Casuso}, E., {Peletier}, R.~F., \& {Beckman}, J.~E. 1996,
  \apjs, 106, 307

\bibitem[{{Wing} \& {Ford}(1969)}]{1969PASP...81..527W}
{Wing}, R.~F., \& {Ford}, Jr., W.~K. 1969, \pasp, 81, 527

\bibitem[{{Zieleniewski} {et~al.}(2015){Zieleniewski}, {Houghton}, {Thatte}, \&
  {Davies}}]{2015MNRAS.452..597Z}
{Zieleniewski}, S., {Houghton}, R.~C.~W., {Thatte}, N., \& {Davies}, R.~L.
  2015, \mnras, 452, 597

\bibitem[{{Zieleniewski} {et~al.}(2017){Zieleniewski}, {Houghton}, {Thatte},
  {Davies}, \& {Vaughan}}]{2017MNRAS.465..192Z}
{Zieleniewski}, S., {Houghton}, R.~C.~W., {Thatte}, N., {Davies}, R.~L., \&
  {Vaughan}, S.~P. 2017, \mnras, 465, 192

\end{thebibliography}
	
	\appendix
	
	\section{Tables of data values}
	
	This Appendix contains the full set of measured equivalent widths for all spectra. 
	
	\vspace{3mm}
	
	Tables A1--A8: measurements for individual galaxies. 
	
	\vspace{1mm}
	
	Table A9: measurements for stacked spectra.
	
	\begin{table*}
		\centering
		
		\caption{Equivalent widths measured from NGC\,0524, 
			given for each radial extraction region and adjusted to a common velocity dispersion of 230\,kms$^{-1}$.
			R1 and R2 correspond to the two extraction regions in the central IFU ($<0.7\arcsec$, $>0.7\arcsec$) 
			while R3, R4, and R5 correspond to the rings of IFUs arranged at 
			$\sim1/3\,$R$_{\rm eff}$, $\sim2/3\,$R$_{\rm eff}$, and $\sim$R$_{\rm eff}$ (N.B. for this galaxy only two rings of IFUs were used).
			Also shown: best-fit velocity dispersion used to correct the measurements, average continuum signal-to-noise ratio in the IZ and YJ band observations.}
		\begin{tabular}{c|ccccc}
			\hline
			NGC\,0524                     &       R1        &       R2        &          R3           &        R4        &       R5        \\ \hline
			Na$\,$I 0.82$\umu$m                & 0.47 $\pm$ 0.14 & 0.17 $\pm$ 0.08 &   --0.04 $\pm$ 0.42   & 0.63 $\pm$ 0.66  & \gap{}---\gap{} \\
			Ca$\,$II Triplet                 & 7.60 $\pm$ 0.25 & 6.95 $\pm$ 0.19 & \gap{}6.57 $\pm$ 0.49 & 6.65 $\pm$ 1.11  & \gap{}---\gap{} \\
			Mg$\,$I 0.88$\umu$m                & 0.59 $\pm$ 0.05 & 0.49 $\pm$ 0.03 & \gap{}0.16 $\pm$ 0.10 & 0.72 $\pm$ 0.26  & \gap{}---\gap{} \\
			Wing-Ford band                  & 0.40 $\pm$ 0.04 & 0.29 $\pm$ 0.03 & \gap{}0.60 $\pm$ 0.10 & 0.37 $\pm$ 0.22  & \gap{}---\gap{} \\
			Ca$\,$I 1.03$\umu$m                & 0.25 $\pm$ 0.06 & 0.45 $\pm$ 0.04 & \gap{}0.41 $\pm$ 0.15 & 0.52 $\pm$ 0.95  & \gap{}---\gap{} \\
			Na$\,$I 1.14$\umu$m                & 1.08 $\pm$ 0.07 & 0.97 $\pm$ 0.06 & \gap{}0.8 $\pm$ 0.12  & 1.03 $\pm$ 0.44  & \gap{}---\gap{} \\
			K$\,$I 1.17$\umu$m a+b              & 0.81 $\pm$ 0.05 & 0.78 $\pm$ 0.03 & \gap{}0.65 $\pm$ 0.09 & 0.75 $\pm$ 0.32  & \gap{}---\gap{} \\
			K$\,$I 1.25$\umu$m                & 0.35 $\pm$ 0.07 & 0.30 $\pm$ 0.06 & \gap{}0.43 $\pm$ 0.10 & 0.73 $\pm$ 0.31  & \gap{}---\gap{} \\
			Al$\,$I 1.31$\umu$m                & 1.42 $\pm$ 0.09 & 1.16 $\pm$ 0.05 & \gap{}1.42 $\pm$ 0.14 & 1.81 $\pm$ 0.64  & \gap{}---\gap{} \\ \hline
			$\sigma_{\mathrm{pPXF}}$ /kms$^{-1}$ (corrected) & 261.0 $\pm$ 3.9 & 217.0 $\pm$ 2.2 &    198.0 $\pm$ 8.6    & 198.0 $\pm$ 17.3 & \gap{}---\gap{} \\
			S/N (IZ)                     &       170       &       254       &          68           &        30        & \gap{}---\gap{} \\
			S/N (YJ)                     &       301       &       450       &          126          &        41        & \gap{}---\gap{} \\ \hline
		\end{tabular}
	\end{table*}
	
	\begin{table*}
		\centering
		
		\caption{Equivalent widths measured from NGC\,1407, 
			given for each radial extraction region and adjusted to a common velocity dispersion of 230\,kms$^{-1}$.
			R1 and R2 correspond to the two extraction regions in the central IFU ($<0.7\arcsec$, $>0.7\arcsec$) 
			while R3, R4, and R5 correspond to the rings of IFUs arranged at 
			$\sim1/3\,$R$_{\rm eff}$, $\sim2/3\,$R$_{\rm eff}$, and $\sim$R$_{\rm eff}$.
			Also shown: best-fit velocity dispersion used to correct the measurements, average continuum signal-to-noise ratio in the IZ and YJ band observations.}
		\begin{tabular}{c|ccccc}
			\hline
			NGC\,1407                     &       R1        &       R2        &          R3           &          R4           &          R5           \\ \hline
			Na$\,$I 0.82$\umu$m                & 0.46 $\pm$ 0.23 & 0.69 $\pm$ 0.10 &   --0.79 $\pm$ 0.22   &   --2.25 $\pm$ 1.00   &   --4.58 $\pm$ 3.30   \\
			Ca$\,$II Triplet                 & 8.08 $\pm$ 0.37 & 7.17 $\pm$ 0.29 & \gap{}6.34 $\pm$ 0.34 & \gap{}8.51 $\pm$ 0.73 & \gap{}7.51 $\pm$ 5.10 \\
			Mg$\,$I 0.88$\umu$m                & 0.60 $\pm$ 0.12 & 0.63 $\pm$ 0.05 & \gap{}0.60 $\pm$ 0.07 & \gap{}0.45 $\pm$ 0.17 & \gap{}0.42 $\pm$ 0.77 \\
			Wing-Ford band                  & 0.02 $\pm$ 0.09 & 0.16 $\pm$ 0.06 & \gap{}0.40 $\pm$ 0.11 & \gap{}0.27 $\pm$ 0.23 & \gap{}0.76 $\pm$ 1.31 \\
			Ca$\,$I 1.03$\umu$m                &       ---       &       ---       &          ---          &          ---          &          ---          \\
			Na$\,$I 1.14$\umu$m                &       ---       &       ---       &          ---          &          ---          &          ---          \\
			K$\,$I 1.17$\umu$m a+b              &       ---       &       ---       &          ---          &          ---          &          ---          \\
			K$\,$I 1.25$\umu$m                &       ---       &       ---       &          ---          &          ---          &          ---          \\
			Al$\,$I 1.31$\umu$m                &       ---       &       ---       &          ---          &          ---          &          ---          \\ \hline
			$\sigma_{\mathrm{pPXF}}$ /kms$^{-1}$ (corrected) & 313.0 $\pm$ 6.6 & 323.0 $\pm$ 6.1 &    234.0 $\pm$ 7.9    &   268.0 $\pm$ 22.6    &   123.0 $\pm$ 41.4    \\
			S/N (IZ)                     &       99        &       168       &          84           &          36           &           6           \\
			S/N (YJ)                     &        0        &        0        &           0           &           0           &           0           \\ \hline
		\end{tabular}
	\end{table*}
	
	\begin{table*}
		\centering
		
		\caption{Equivalent widths measured from NGC\,3377, 
			given for each radial extraction region and adjusted to a common velocity dispersion of 230\,kms$^{-1}$.
			R1 and R2 correspond to the two extraction regions in the central IFU ($<0.7\arcsec$, $>0.7\arcsec$) 
			while R3, R4, and R5 correspond to the rings of IFUs arranged at 
			$\sim1/3\,$R$_{\rm eff}$, $\sim2/3\,$R$_{\rm eff}$, and $\sim$R$_{\rm eff}$.
			Also shown: best-fit velocity dispersion used to correct the measurements, average continuum signal-to-noise ratio in the IZ and YJ band observations.}
		\begin{tabular}{c|ccccc}
			\hline
			NGC\,3377                     &       R1        &          R2           &          R3           &          R4           &          R5           \\ \hline
			Na$\,$I 0.82$\umu$m                & 0.29 $\pm$ 0.07 & \gap{}0.34 $\pm$ 0.05 & \gap{}0.07 $\pm$ 0.39 &   --0.13 $\pm$ 0.78   &   --0.66 $\pm$ 1.00   \\
			Ca$\,$II Triplet                 & 5.94 $\pm$ 0.18 & \gap{}7.12 $\pm$ 0.11 & \gap{}6.31 $\pm$ 0.63 & \gap{}2.71 $\pm$ 1.71 & \gap{}2.00 $\pm$ 2.16 \\
			Mg$\,$I 0.88$\umu$m                & 0.44 $\pm$ 0.04 & \gap{}0.49 $\pm$ 0.02 & \gap{}0.38 $\pm$ 0.11 & \gap{}0.24 $\pm$ 0.17 & \gap{}0.09 $\pm$ 0.42 \\
			Wing-Ford band                  & 0.25 $\pm$ 0.04 & \gap{}0.21 $\pm$ 0.04 & \gap{}0.06 $\pm$ 0.18 &   --0.72 $\pm$ 0.27   &   --0.11 $\pm$ 0.74   \\
			Ca$\,$I 1.03$\umu$m                & 0.27 $\pm$ 0.08 &   --0.22 $\pm$ 0.07   &   --0.14 $\pm$ 0.25   &   --1.24 $\pm$ 0.64   &   --2.06 $\pm$ 1.06   \\
			Na$\,$I 1.14$\umu$m                & 1.54 $\pm$ 0.09 & \gap{}1.04 $\pm$ 0.09 & \gap{}0.41 $\pm$ 1.42 & \gap{}1.35 $\pm$ 1.93 & \gap{}0.41 $\pm$ 2.06 \\
			K$\,$I 1.17$\umu$m a+b              & 0.75 $\pm$ 0.05 & \gap{}0.82 $\pm$ 0.05 & \gap{}0.37 $\pm$ 0.51 & \gap{}1.21 $\pm$ 0.78 & \gap{}0.88 $\pm$ 0.88 \\
			K$\,$I 1.25$\umu$m                & 0.13 $\pm$ 0.07 & \gap{}0.22 $\pm$ 0.07 & \gap{}0.25 $\pm$ 0.34 & \gap{}1.65 $\pm$ 0.74 & \gap{}1.28 $\pm$ 1.58 \\
			Al$\,$I 1.31$\umu$m                & 1.07 $\pm$ 0.08 & \gap{}1.27 $\pm$ 0.08 & \gap{}0.46 $\pm$ 1.1  &   --1.09 $\pm$ 1.78   &   --1.14 $\pm$ 3.09   \\ \hline
			$\sigma_{\mathrm{pPXF}}$ /kms$^{-1}$ (corrected) & 142.0 $\pm$ 1.1 &    104.0 $\pm$ 1.4    &    149.0 $\pm$ 8.2    &    99.0 $\pm$ 15.5    &    94.0 $\pm$ 4.0     \\
			S/N (IZ)                     &       211       &          243          &          52           &          29           &          11           \\
			S/N (YJ)                     &       191       &          236          &          27           &          12           &          11           \\ \hline
		\end{tabular}
	\end{table*}
	
	\begin{table*}
		\centering
		
		\caption{Equivalent widths measured from NGC\,3379, 
			given for each radial extraction region and adjusted to a common velocity dispersion of 230\,kms$^{-1}$.
			R1 and R2 correspond to the two extraction regions in the central IFU ($<0.7\arcsec$, $>0.7\arcsec$) 
			while R3, R4, and R5 correspond to the rings of IFUs arranged at 
			$\sim1/3\,$R$_{\rm eff}$, $\sim2/3\,$R$_{\rm eff}$, and $\sim$R$_{\rm eff}$.
			Also shown: best-fit velocity dispersion used to correct the measurements, average continuum signal-to-noise ratio in the IZ and YJ band observations.}
		\begin{tabular}{c|ccccc}
			\hline
			NGC\,3379                     &       R1        &       R2        &       R3        &       R4        &          R5           \\ \hline
			Na$\,$I 0.82$\umu$m                & 0.26 $\pm$ 0.06 & 0.26 $\pm$ 0.03 & 1.02 $\pm$ 0.30 & 0.77 $\pm$ 0.28 &   --0.24 $\pm$ 4.51   \\
			Ca$\,$II Triplet                 & 6.65 $\pm$ 0.15 & 6.35 $\pm$ 0.15 & 6.90 $\pm$ 0.30 & 5.30 $\pm$ 0.41 & \gap{}7.54 $\pm$ 5.97 \\
			Mg$\,$I 0.88$\umu$m                & 0.58 $\pm$ 0.07 & 0.49 $\pm$ 0.04 & 0.48 $\pm$ 0.06 & 0.31 $\pm$ 0.08 & \gap{}0.53 $\pm$ 0.84 \\
			Wing-Ford band                  & 0.10 $\pm$ 0.05 & 0.12 $\pm$ 0.04 & 0.11 $\pm$ 0.07 & 0.83 $\pm$ 0.15 &   --0.08 $\pm$ 1.6    \\
			Ca$\,$I 1.03$\umu$m                &       ---       &       ---       &       ---       &       ---       &          ---          \\
			Na$\,$I 1.14$\umu$m                &       ---       &       ---       &       ---       &       ---       &          ---          \\
			K$\,$I 1.17$\umu$m a+b              &       ---       &       ---       &       ---       &       ---       &          ---          \\
			K$\,$I 1.25$\umu$m                &       ---       &       ---       &       ---       &       ---       &          ---          \\
			Al$\,$I 1.31$\umu$m                &       ---       &       ---       &       ---       &       ---       &          ---          \\ \hline
			$\sigma_{\mathrm{pPXF}}$ /kms$^{-1}$ (corrected) & 226.0 $\pm$ 1.5 & 205.0 $\pm$ 1.4 & 160.0 $\pm$ 3.1 & 108.0 $\pm$ 6.3 &    97.0 $\pm$ 90.3    \\
			S/N (IZ)                     &       238       &       394       &       120       &       62        &           5           \\
			S/N (YJ)                     &        0        &        0        &        0        &        0        &           0           \\ \hline
		\end{tabular}
	\end{table*}
	
	\begin{table*}
		\centering
		
		\caption{Equivalent widths measured from NGC\,4486, 
			given for each radial extraction region and adjusted to a common velocity dispersion of 230\,kms$^{-1}$.
			R1 and R2 correspond to the two extraction regions in the central IFU ($<0.7\arcsec$, $>0.7\arcsec$) 
			while R3, R4, and R5 correspond to the rings of IFUs arranged at 
			$\sim1/3\,$R$_{\rm eff}$, $\sim2/3\,$R$_{\rm eff}$, and $\sim$R$_{\rm eff}$.
			Also shown: best-fit velocity dispersion used to correct the measurements, average continuum signal-to-noise ratio in the IZ and YJ band observations.}
		\begin{tabular}{c|ccccc}
			\hline
			NGC\,4486                     &        R1        &       R2        &          R3           &          R4           &          R5           \\ \hline
			Na$\,$I 0.82$\umu$m                & 0.47 $\pm$ 0.45  & 1.34 $\pm$ 0.19 & \gap{}1.28 $\pm$ 0.30 &   --0.02 $\pm$ 0.61   &   --0.87 $\pm$ 1.93   \\
			Ca$\,$II Triplet                 & 8.29 $\pm$ 0.62  & 5.93 $\pm$ 0.23 & \gap{}5.78 $\pm$ 0.46 & \gap{}5.67 $\pm$ 0.66 & \gap{}3.45 $\pm$ 2.21 \\
			Mg$\,$I 0.88$\umu$m                & 0.47 $\pm$ 0.05  & 0.60 $\pm$ 0.05 & \gap{}0.59 $\pm$ 0.09 & \gap{}0.29 $\pm$ 0.16 & \gap{}0.40 $\pm$ 0.91 \\
			Wing-Ford band                  &       ---        &       ---       & \gap{}0.24 $\pm$ 0.14 & \gap{}0.74 $\pm$ 0.25 & \gap{}0.65 $\pm$ 2.31 \\
			Ca$\,$I 1.03$\umu$m                &       ---        &       ---       & \gap{}0.08 $\pm$ 0.17 & \gap{}0.02 $\pm$ 0.31 &   --0.27 $\pm$ 2.71   \\
			Na$\,$I 1.14$\umu$m                & 2.65 $\pm$ 0.26  & 2.16 $\pm$ 0.12 & \gap{}1.83 $\pm$ 0.20 & \gap{}1.69 $\pm$ 0.82 &   --1.05 $\pm$ 1.94   \\
			K$\,$I 1.17$\umu$m a+b              & 0.87 $\pm$ 0.19  & 0.78 $\pm$ 0.08 & \gap{}1.16 $\pm$ 0.10 & \gap{}0.39 $\pm$ 0.31 & \gap{}1.31 $\pm$ 0.72 \\
			K$\,$I 1.25$\umu$m                &       ---        &       ---       &   --0.28 $\pm$ 0.12   & \gap{}0.72 $\pm$ 0.26 & \gap{}0.18 $\pm$ 0.55 \\
			Al$\,$I 1.31$\umu$m                & 1.25 $\pm$ 0.23  & 1.16 $\pm$ 0.07 & \gap{}1.30 $\pm$ 0.12 & \gap{}0.45 $\pm$ 0.34 &   --0.84 $\pm$ 0.99   \\ \hline
			$\sigma_{\mathrm{pPXF}}$ /kms$^{-1}$ (corrected) & 400.0 $\pm$ 11.8 & 341.0 $\pm$ 2.9 &    293.0 $\pm$ 4.7    &   257.0 $\pm$ 11.2    &   300.0 $\pm$ 22.7    \\
			S/N (IZ)                     &       108        &       165       &          81           &          45           &          11           \\
			S/N (YJ)                     &       164        &       299       &          144          &          50           &          23           \\ \hline
		\end{tabular}
	\end{table*}

	\begin{table*}
		\centering
		
		\caption{Equivalent widths measured from NGC\,4552, 
			given for each radial extraction region and adjusted to a common velocity dispersion of 230\,kms$^{-1}$.
			R1 and R2 correspond to the two extraction regions in the central IFU ($<0.7\arcsec$, $>0.7\arcsec$) 
			while R3, R4, and R5 correspond to the rings of IFUs arranged at 
			$\sim1/3\,$R$_{\rm eff}$, $\sim2/3\,$R$_{\rm eff}$, and $\sim$R$_{\rm eff}$.
			Also shown: best-fit velocity dispersion used to correct the measurements, average continuum signal-to-noise ratio in the IZ and YJ band observations.}
		\begin{tabular}{c|ccccc}
			\hline
			NGC\,4552                     &       R1        &       R2        &       R3        &          R4           &          R5           \\ \hline
			Na$\,$I 0.82$\umu$m                & 1.03 $\pm$ 0.13 & 0.75 $\pm$ 0.06 & 0.46 $\pm$ 0.23 & \gap{}0.19 $\pm$ 1.00 & \gap{}1.45 $\pm$ 0.69 \\
			Ca$\,$II Triplet                 & 6.41 $\pm$ 0.22 & 6.29 $\pm$ 0.16 & 6.48 $\pm$ 0.57 & \gap{}4.40 $\pm$ 1.53 & \gap{}6.08 $\pm$ 1.50 \\
			Mg$\,$I 0.88$\umu$m                & 0.72 $\pm$ 0.06 & 0.48 $\pm$ 0.04 & 0.36 $\pm$ 0.10 & \gap{}0.19 $\pm$ 0.21 & \gap{}0.29 $\pm$ 0.24 \\
			Wing-Ford band                  & 0.22 $\pm$ 0.07 & 0.21 $\pm$ 0.05 & 0.16 $\pm$ 0.11 &   --0.34 $\pm$ 0.31   & \gap{}0.92 $\pm$ 0.66 \\
			Ca$\,$I 1.03$\umu$m                & 0.28 $\pm$ 0.09 & 0.22 $\pm$ 0.05 & 0.34 $\pm$ 0.13 & \gap{}0.49 $\pm$ 0.33 & \gap{}0.27 $\pm$ 0.74 \\
			Na$\,$I 1.14$\umu$m                & 2.24 $\pm$ 0.08 & 1.95 $\pm$ 0.06 & 1.79 $\pm$ 0.20 & \gap{}1.33 $\pm$ 0.98 & \gap{}1.83 $\pm$ 0.74 \\
			K$\,$I 1.17$\umu$m a+b              & 0.77 $\pm$ 0.07 & 0.82 $\pm$ 0.06 & 0.53 $\pm$ 0.18 & \gap{}0.18 $\pm$ 1.32 & \gap{}0.71 $\pm$ 0.63 \\
			K$\,$I 1.25$\umu$m                & 0.33 $\pm$ 0.06 & 0.22 $\pm$ 0.05 & 0.06 $\pm$ 0.13 &   --0.35 $\pm$ 0.85   & \gap{}0.04 $\pm$ 0.56 \\
			Al$\,$I 1.31$\umu$m                & 1.52 $\pm$ 0.08 & 1.49 $\pm$ 0.06 & 0.48 $\pm$ 0.36 &   --0.49 $\pm$ 3.08   &   --1.95 $\pm$ 1.22   \\ \hline
			$\sigma_{\mathrm{pPXF}}$ /kms$^{-1}$ (corrected) & 256.0 $\pm$ 1.6 & 234.0 $\pm$ 1.2 & 178.0 $\pm$ 4.2 &    160.0 $\pm$ 9.1    &   124.0 $\pm$ 10.5    \\
			S/N (IZ)                     &       154       &       244       &       75        &          28           &          18           \\
			S/N (YJ)                     &       226       &       343       &       70        &          11           &          18           \\ \hline
		\end{tabular}
	\end{table*}
	
	\begin{table*}
		\centering
		
		\caption{Equivalent widths measured from NGC\,4621, 
			given for each radial extraction region and adjusted to a common velocity dispersion of 230\,kms$^{-1}$.
			R1 and R2 correspond to the two extraction regions in the central IFU ($<0.7\arcsec$, $>0.7\arcsec$) 
			while R3, R4, and R5 correspond to the rings of IFUs arranged at 
			$\sim1/3\,$R$_{\rm eff}$, $\sim2/3\,$R$_{\rm eff}$, and $\sim$R$_{\rm eff}$.
			Also shown: best-fit velocity dispersion used to correct the measurements, average continuum signal-to-noise ratio in the IZ and YJ band observations.}
		\begin{tabular}{c|ccccc}
			\hline
			NGC\,4621                     &       R1        &       R2        &          R3           &       R4        &          R5           \\ \hline
			Na$\,$I 0.82$\umu$m                & 1.18 $\pm$ 0.10 & 0.70 $\pm$ 0.05 & \gap{}0.36 $\pm$ 0.20 & 0.09 $\pm$ 0.36 &   --0.74 $\pm$ 0.37   \\
			Ca$\,$II Triplet                 & 7.26 $\pm$ 0.21 & 7.03 $\pm$ 0.14 & \gap{}6.04 $\pm$ 0.42 & 7.08 $\pm$ 0.68 & \gap{}3.87 $\pm$ 0.73 \\
			Mg$\,$I 0.88$\umu$m                & 0.70 $\pm$ 0.06 & 0.62 $\pm$ 0.04 & \gap{}0.59 $\pm$ 0.10 & 0.44 $\pm$ 0.15 & \gap{}0.72 $\pm$ 0.21 \\
			Wing-Ford band                  & 0.23 $\pm$ 0.04 & 0.21 $\pm$ 0.03 & \gap{}0.37 $\pm$ 0.10 & 0.39 $\pm$ 0.19 &   --0.12 $\pm$ 0.29   \\
			Ca$\,$I 1.03$\umu$m                & 0.56 $\pm$ 0.05 & 0.56 $\pm$ 0.04 & \gap{}0.68 $\pm$ 0.14 & 0.44 $\pm$ 0.20 &   --0.15 $\pm$ 0.33   \\
			Na$\,$I 1.14$\umu$m                & 1.69 $\pm$ 0.04 & 1.50 $\pm$ 0.04 & \gap{}1.03 $\pm$ 0.16 & 0.70 $\pm$ 0.20 & \gap{}2.12 $\pm$ 0.39 \\
			K$\,$I 1.17$\umu$m a+b              & 0.98 $\pm$ 0.06 & 0.90 $\pm$ 0.04 & \gap{}0.97 $\pm$ 0.14 & 0.88 $\pm$ 0.25 & \gap{}1.71 $\pm$ 0.55 \\
			K$\,$I 1.25$\umu$m                & 0.07 $\pm$ 0.06 & 0.05 $\pm$ 0.06 &   --0.06 $\pm$ 0.15   & 0.12 $\pm$ 0.35 & \gap{}1.12 $\pm$ 0.59 \\
			Al$\,$I 1.31$\umu$m                & 1.71 $\pm$ 0.07 & 1.59 $\pm$ 0.05 & \gap{}1.38 $\pm$ 0.16 & 0.79 $\pm$ 0.27 &   --1.21 $\pm$ 0.93   \\ \hline
			$\sigma_{\mathrm{pPXF}}$ /kms$^{-1}$ (corrected) & 270.0 $\pm$ 1.3 & 228.0 $\pm$ 1.3 &    207.0 $\pm$ 3.3    & 192.0 $\pm$ 7.7 &   217.0 $\pm$ 17.9    \\
			S/N (IZ)                     &       223       &       286       &          93           &       52        &          40           \\
			S/N (YJ)                     &       343       &       480       &          141          &       75        &          32           \\ \hline
		\end{tabular}
	\end{table*}
	
	\begin{table*}
		\centering
		
		\caption{Equivalent widths measured from NGC\,5813, 
			given for each radial extraction region and adjusted to a common velocity dispersion of 230\,kms$^{-1}$.
			R1 and R2 correspond to the two extraction regions in the central IFU ($<0.7\arcsec$, $>0.7\arcsec$) 
			while R3, R4, and R5 correspond to the rings of IFUs arranged at 
			$\sim1/3\,$R$_{\rm eff}$, $\sim2/3\,$R$_{\rm eff}$, and $\sim$R$_{\rm eff}$.
			Also shown: best-fit velocity dispersion used to correct the measurements, average continuum signal-to-noise ratio in the IZ and YJ band observations.}
		\begin{tabular}{c|ccccc}
			\hline
			NGC\,5813                     &          R1           &          R2           &          R3           &          R4           &          R5           \\ \hline
			Na$\,$I 0.82$\umu$m                & \gap{}0.15 $\pm$ 0.14 & \gap{}0.33 $\pm$ 0.09 &   --0.49 $\pm$ 0.88   &   --0.59 $\pm$ 0.83   &   --0.43 $\pm$ 1.53   \\
			Ca$\,$II Triplet                 & \gap{}6.64 $\pm$ 0.27 & \gap{}7.10 $\pm$ 0.20 & \gap{}5.62 $\pm$ 1.13 & \gap{}8.20 $\pm$ 1.51 & \gap{}6.15 $\pm$ 4.64 \\
			Mg$\,$I 0.88$\umu$m                & \gap{}0.49 $\pm$ 0.06 & \gap{}0.44 $\pm$ 0.05 & \gap{}0.42 $\pm$ 0.18 & \gap{}0.73 $\pm$ 0.22 & \gap{}0.54 $\pm$ 0.66 \\
			Wing-Ford band                  & \gap{}0.18 $\pm$ 0.08 & \gap{}0.36 $\pm$ 0.06 & \gap{}0.29 $\pm$ 0.25 & \gap{}1.04 $\pm$ 0.38 & \gap{}0.75 $\pm$ 1.42 \\
			Ca$\,$I 1.03$\umu$m                & \gap{}0.41 $\pm$ 0.09 & \gap{}0.48 $\pm$ 0.07 & \gap{}0.43 $\pm$ 1.24 &   --0.63 $\pm$ 0.58   & \gap{}0.93 $\pm$ 1.50 \\
			Na$\,$I 1.14$\umu$m                & \gap{}1.25 $\pm$ 0.09 & \gap{}1.29 $\pm$ 0.07 & \gap{}1.35 $\pm$ 0.89 &   --0.07 $\pm$ 0.43   &   --1.63 $\pm$ 1.25   \\
			K$\,$I 1.17$\umu$m a+b              & \gap{}0.73 $\pm$ 0.07 & \gap{}0.64 $\pm$ 0.05 & \gap{}0.50 $\pm$ 0.48 & \gap{}0.61 $\pm$ 0.25 & \gap{}0.93 $\pm$ 0.60 \\
			K$\,$I 1.25$\umu$m                &   --0.12 $\pm$ 0.09   &   --0.02 $\pm$ 0.07   &   --0.02 $\pm$ 0.57   &   --0.43 $\pm$ 0.4    &   --0.88 $\pm$ 0.86   \\
			Al$\,$I 1.31$\umu$m                & \gap{}0.56 $\pm$ 0.08 & \gap{}0.63 $\pm$ 0.06 & \gap{}0.71 $\pm$ 1.32 &   --0.23 $\pm$ 0.49   &   --0.86 $\pm$ 1.56   \\ \hline
			$\sigma_{\mathrm{pPXF}}$ /kms$^{-1}$ (corrected) &    249.0 $\pm$ 2.8    &    221.0 $\pm$ 2.8    &   219.0 $\pm$ 14.9    &    119.0 $\pm$ 1.9    &   133.0 $\pm$ 30.2    \\
			S/N (IZ)                     &          135          &          182          &          31           &          19           &           7           \\
			S/N (YJ)                     &          230          &          294          &          20           &          35           &          14           \\ \hline
		\end{tabular}
	\end{table*}
	
	\begin{table*}
		\centering
		\caption{Equivalent widths measured from the stacked spectra, given for each radial extraction region 
			and adjusted to common velocity dispersion of 230\,kms$^{-1}$. Definitions as in Table A1.}
		\begin{tabular}{c|ccccc}
			\hline
			Index          &       R1       &       R2       &       R3       &       R4       &          R5          \\ \hline
			Na$\,$I 0.82$\umu$m   & 0.38$\pm$ 0.20 & 0.38$\pm$ 0.09 & 0.08$\pm$ 0.19 & 0.16$\pm$ 0.22 &   --0.5$\pm$ 0.43    \\
			Ca$\,$II Triplet    & 6.62$\pm$ 0.25 & 6.38$\pm$ 0.19 & 6.23$\pm$ 0.25 & 5.91$\pm$ 0.50 & \gap{}5.33$\pm$ 0.69 \\
			Mg$\,$I 0.88$\umu$m   & 0.51$\pm$ 0.05 & 0.50$\pm$ 0.03 & 0.39$\pm$ 0.06 & 0.40$\pm$ 0.10 & \gap{}0.51$\pm$ 0.14 \\
			Wing-Ford band     & 0.21$\pm$ 0.05 & 0.24$\pm$ 0.03 & 0.32$\pm$ 0.06 & 0.30$\pm$ 0.18 & \gap{}0.29$\pm$ 0.24 \\
			Ca$\,$I 1.03$\umu$m   & 0.31$\pm$ 0.03 & 0.36$\pm$ 0.03 & 0.47$\pm$ 0.08 & 0.25$\pm$ 0.07 & \gap{}0.32$\pm$ 0.22 \\
			Na$\,$I 1.14$\umu$m   & 1.56$\pm$ 0.24 & 1.32$\pm$ 0.09 & 1.05$\pm$ 0.14 & 0.98$\pm$ 0.25 & \gap{}1.18$\pm$ 0.78 \\
			K$\,$I 1.17$\umu$m a+b & 0.75$\pm$ 0.07 & 0.73$\pm$ 0.04 & 0.62$\pm$ 0.10 & 0.67$\pm$ 0.11 & \gap{}1.07$\pm$ 0.20 \\
			K$\,$I 1.25$\umu$m   & 0.08$\pm$ 0.15 & 0.05$\pm$ 0.08 & 0.06$\pm$ 0.14 & 0.23$\pm$ 0.28 & \gap{}0.34$\pm$ 0.86 \\
			Al$\,$I 1.31$\umu$m   & 1.26$\pm$ 0.23 & 1.24$\pm$ 0.09 & 1.15$\pm$ 0.13 & 0.83$\pm$ 0.34 & \gap{}0.41$\pm$ 0.69 \\ \hline
		\end{tabular}
	\end{table*}
	
	\section{Correction for AGN emission}
	
	Before interpreting the measured index equivalent widths, we assessed the impact of contamination of our absorption features by active galactic nuclei (AGN). Four of our sample are known to contain low-ionisation nuclear emission regions (LINERs): NGC$\,$3379, NGC$\,$4486, NGC$\,$4552, and NGC$\,$5813 (see for example \cite{1997ApJS..112..315H}). In particular, strong nuclear emission lines were observed in the core spectrum of NGC$\,$4486 when the region within a 0.7$\arcsec$ radius was isolated. To assess how widespread this contamination might be, we created maps of the strengths of two emission lines: the [S$\,$III] $\lambda\,$9533 $+$ Pa-$\epsilon$ composite line, and various other lines known to be strong in many AGN -- see \cite{2008ApJS..174..282L}. For this purpose, the data from the central IFU in each galaxy was re-reduced, using a straightforward subtraction of the datacubes in each pair of exposures (justifiable due to the high signal-to-noise of these cubes) and applying the telluric correction derived in the full reduction pipeline. The three exposures were then combined, taking offsets in the targeting into account. This method preserved the full spatial information carried in the datacubes.
	
	\begin{figure*}
		\centering
		\includegraphics[width=0.99\textwidth]{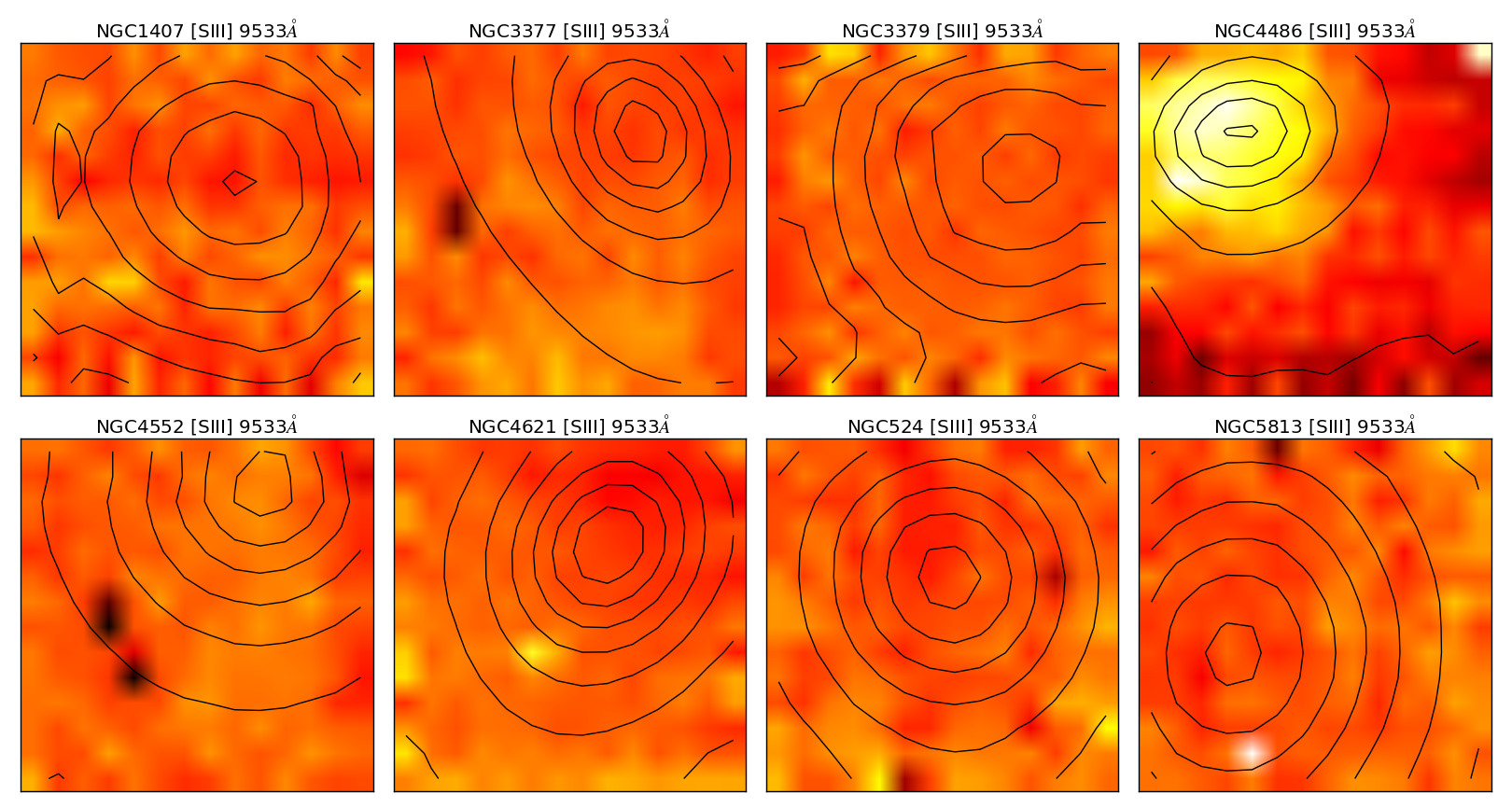}
		\caption{Maps of the strength of [S$\,$III] ($\lambda\,$9533) emission in a central $2.8\arcsec \times 2.8\arcsec$ IFU arm for each galaxy (normalised by the continuum emission level). Pixel coordinates. Black contours indicate continuum emission derived by collapsing the datacube along the wavelength axis.}
	\end{figure*}
	
	It is clear from Fig. B1 (in which we show the strongest emission line signature we identified) that only in NGC$\,$4486 is AGN line emission detectable, and even then it is only significant in the very centre. Although most of our absorption features in this spectrum are not compromised by AGN emission lines, continuum AGN emission could contaminate all the absorption features in that spectrum (an additional continuum component will reduce the equivalent width measurement for each index). 
	
	We estimate the level of AGN continuum contamination as follows: In \cite{2016MNRAS.457.3801P}, Hubble observations of the central parsecs of M87 are used to measure the AGN flux at wavelengths of $\sim1\,\umu$m. This is quoted as $\approx10^{-3}\,$Jy, or magnitude 15.5. Meanwhile, the 2MASS large galaxies atlas \citep{2003AJ....125..525J}, quotes the core surface brightness (within 2$\arcsec$ of the centre) as 14.7 mag arcsec$^{-2}$. 
	
	From this, we calculate the total flux expected within the R$<0.7\arcsec$ region, which will contain almost all the AGN flux, assuming a de Vaucouleurs brightness profile. We obtain a magnitude of 13.4 for the extraction region, which we compare with the AGN magnitude of 15.5 to estimate an AGN contribution of $\sim15\%$ to the total continuum level.

	This estimate allowed us to make a rough correction to the equivalent widths measured for absorption lines in NGC\,4486's central spectrum. The small number of feature measurements contaminated by line emission (the Wing-Ford band, Ca$\,$I at 1.03$\,\umu$m, and K$\,$I at 1.25$\,\umu$m -- contaminated by [C$\,$I] $\lambda\,$9853, [S$\,$II] $\lambda\,$10323, and [Fe$\,$II] $\lambda\,$12570 respectively) in this spectrum were not included in the analysis.
	
	\section{Parametrizing IMF variations}
	
	Certain spectral features in the integrated light of galaxies are sensitive to the IMF. This is because there are some stellar spectral features which are `gravity sensitive'. These features have different strengths in dwarf and giant stars with the same effective temperature (blackbody spectral shape) and chemical composition, due to the variation of the detailed atmospheric physics under changes in surface gravity. The strength of these features in the \textit{integrated light} depends on the relative contribution of the dwarfs and the giants to the total light. This is determined by the IMF and the age of the stellar population.
	
	We have a set of SSP models computed at solar metallicity and old (t=13.5\,Gyr) age for a variety of IMFs. The variation of these models over a sensible range of ages is slight in our wavelength range, so we can drop the fixed age requirement and look at the strength of our set of absorption indices for a range of IMFs. We seek a parameter that encapsulates the variation in the IMF, against which the variation in index strength is approximately linear. The best choice is $f_{\mathrm{dwarf}}$, the fraction of the light contributed by stars with mass $<$ M$_{\rm dwarf}$, since this is precisely the quantity on which the index strengths depend.
	
	We compute $f_{\mathrm{dwarf}} = \mathrm{ \frac{1}{L_{tot}} \int_{0.1M_{\odot}}^{M_{dwarf}} L(M;t_{pop})\,N(M)\,dM }$ for a variety of choices of M$_{\rm dwarf}$. To do this we combine the IMF functional forms N(M) used to create the CvD SSP models with the BaSTI isochrones L(M; t$_{\rm pop}$) computed at t=13.5\,Gyr and 10.0\,Gyr. These isochrones are only computed down to 0.1\,M$_{\odot}$, rather than 0.08\,M$_{\odot}$ (which is what the SSP models use as the low mass cutoff) but the contribution of the sub-0.1\,M$_{\odot}$ dwarfs to the total light is marginal and should not change our calculations significantly. The contributions to the light would be smaller for a younger population (slightly higher mass stars are still extant), but since the index strengths don't significantly change this doesn't matter, since we will just be converting any inferred change in $f_{\mathrm{dwarf}}$ back to a change in the IMF \textit{via the same assumed age}. In any case, between e.g. a 10\,Gyr and 13.5\,Gyr old stellar population, the change in the slope of the relation between index strength and $f_{\mathrm{dwarf}}$ is of order a few percent.
	
	In Tables C1 and C2 we show $f_{\mathrm{dwarf}}$ for several IMFs and choices of M$_{\mathrm{dwarf}}$. We can compare IMFs described with a single power-law with Vazdekis-type IMFs (in which the power law breaks at 0.6M$_{\odot}$ and is shallower at low mass -- see \citealp{1996ApJS..106..307V}) and the Chabrier (Milky-Way) IMF. In figure C1 we show the linearity of the model index strengths in f$_{\rm dwarf}$ with index strengths shown for the Chabrier IMF and three power-law IMFs (X=2.3, 3.0, and 3.5). Particular attention should be paid to those indices with strong variation against the IMF. We note that in these indices, the detailed shape of the IMF seems to be a second-order effect (\textit{viz.} the typically small deviation of the models generated with a Chabrier IMF from the linear relation given by the models generated using three single-slope power law IMFs), and that the model index strengths do seem to follow a linear relation.
	
	For these IMFs the choice of M$_{\rm dwarf}$ does not particularly matter. However, for Vazdekis IMFs the contribution to the light of the lowest mass stars does not change much with the slope. This suggests that our formalism ought to be most robust against changes in the IMF functional form if a high value for M$_{\rm dwarf}$ is used. We therefore use M$_{\rm dwarf} = 0.5M_{\odot}$ throughout this work.
	
	\begin{table*}
		\centering
		
		\caption{$f_{\mathrm{dwarf}}$ is the fraction of the continuum J-band luminosity provided by stars with masses below M$_{\rm dwarf}$. This table gives $f_{\mathrm{dwarf}}$ values for a variety of IMFs and definitions of M$_{\mathrm{dwarf}}$. Note that Vazdekis-type IMFs are usually quoted in logarithmic units, a formalism we use in this table for $\Gamma_{b}$, the slope above 0.6M$_{\odot}$. $\Gamma_{b}$+1 gives X, so in this formalism $\Gamma_{b} = 1.3$ corresponds to the Salpeter slope at high mass. Computed at t=13.5Gyr.}
		\begin{tabular}{c|cccccc}
			\hline
			M$_{\mathrm{dwarf}}$ & Chabrier & X=2.3 & X=3.0 & X=3.5 & Vazdekis ($\Gamma_{b}$ = 1.3) & ($\Gamma_{b}$ = 3.1) \\ \hline
			$<$0.5M$_{\odot}$   &   4.3\%    &  8.4\%  & 19.0\%  & 32.6\%  &              3.6\%              &         7.2\%          \\
			$<$0.4M$_{\odot}$   &   3.0\%    &  6.6\%  & 16.3\%  & 29.4\%  &              2.2\%              &         4.5\%          \\
			$<$0.3M$_{\odot}$   &   1.8\%    &  4.7\%  & 12.9\%  & 24.7\%  &              1.2\%              &         2.4\%          \\
			$<$0.2M$_{\odot}$   &   0.7\%    &  2.4\%  &  7.9\%  & 16.7\%  &              0.4\%              &         0.9\%          \\ \hline
		\end{tabular}
	\end{table*}

	\begin{table*}
		\centering
		
		\caption{$f_{\mathrm{dwarf}}$ is the fraction of the continuum J-band luminosity provided by stars with masses below M$_{\rm dwarf}$. This table gives $f_{\mathrm{dwarf}}$ values for a variety of IMFs and definitions of M$_{\mathrm{dwarf}}$. Computed at t=10.0\,Gyr.}
		\begin{tabular}{c|cccccc}
			\hline
			M$_{\mathrm{dwarf}}$ & Chabrier & X=2.3 & X=3.0  & X=3.5  & Vazdekis ($\Gamma_{b}$ = 1.3) & ($\Gamma_{b}$ = 3.1) \\ \hline
			$<$0.5M$_{\odot}$   &  3.3\%   & 6.6\% & 16.1\% & 29.1\% &             2.8\%             &        6.4\%         \\
			$<$0.4M$_{\odot}$   &  2.3\%   & 5.2\% & 13.8\% & 26.2\% &             1.7\%             &        4.0\%         \\
			$<$0.3M$_{\odot}$   &  1.4\%   & 3.7\% & 10.9\% & 22.1\% &             0.9\%             &        2.1\%         \\
			$<$0.2M$_{\odot}$   &  0.5\%   & 1.9\% & 6.7\%  & 14.9\% &             0.3\%             &        0.8\%         \\ \hline
		\end{tabular}
	\end{table*}

	\begin{figure*}
		\centering
		\includegraphics[width=0.99\textwidth]{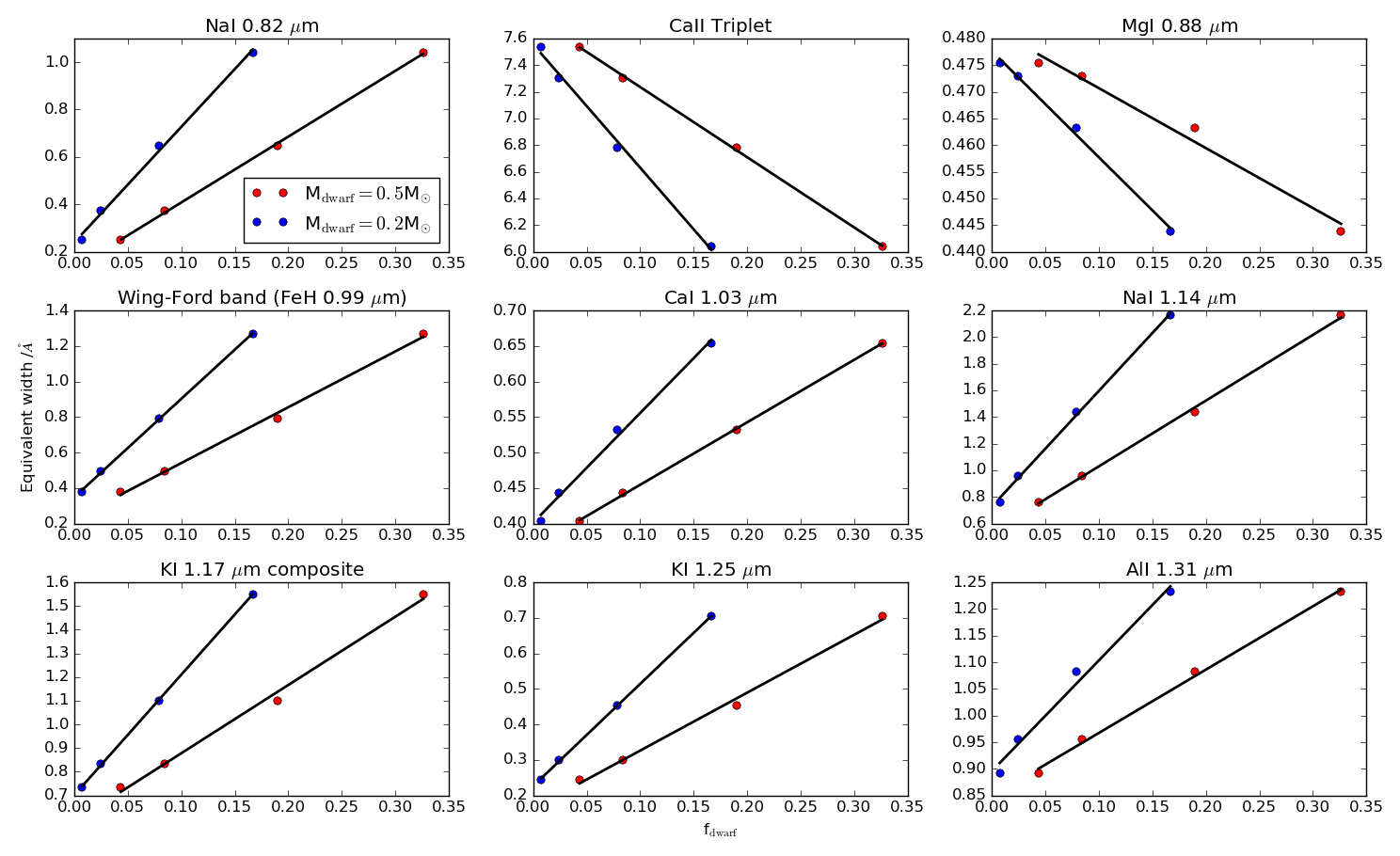}
		\vskip 3mm
		\includegraphics[width=0.99\textwidth]{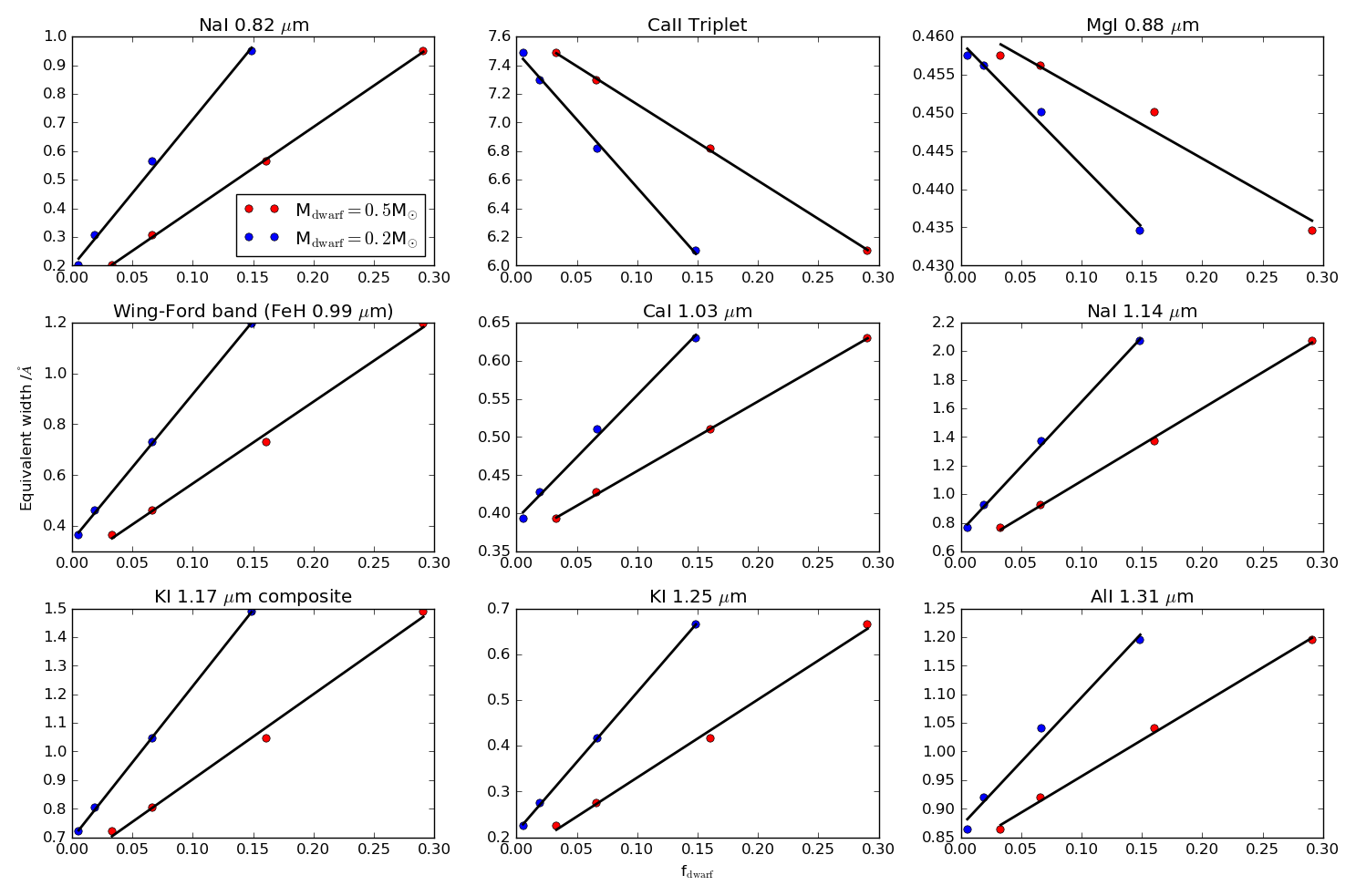}
		\caption{ -- This plot displays the linear variation of each of our indices with f$_{\mathrm{dwarf}}$, for two different definitions of M$_{\mathrm{dwarf}}$, computed at t=13.5\,Gyr (top panel) and t=10.0\,Gyr (bottom panel). The models used to generate the predicted index strengths are generated via a Chabrier IMF and single-slope power laws with slopes X=2.3, 3.0, and 3.5.}
	\end{figure*}
	
	\label{lastpage}

\end{document}